\documentclass{aastex62}

\usepackage{amsmath}
\usepackage{amsfonts}
\usepackage{amssymb}
\def\simlt{\lower.5ex\hbox{$\; \buildrel < \over \sim \;$}}
\def\simgt{\lower.5ex\hbox{$\; \buildrel > \over \sim \;$}}
\def\kms{km s$^{-1}$}

\def\msun{{$M_\odot$}}
\def\lsun{{$L_\odot$}}

\def\mum{$\mu$m}

\def\vlsr{v_{\rm LSR}}

\def\spitzer{{\it Spitzer}}
\def\herschel{{\it Herschel}}

\newcommand{\feii}{[\ion{Fe}{2}]}

\newcommand{\sii}{[\ion{Si}{1}]}
\newcommand{\siii}{[\ion{Si}{2}]}
\newcommand{\oiii}{[\ion{O}{3}]}
\newcommand{\oiv}{[\ion{O}{4}]}
\newcommand{\nii}{[\ion{N}{2}]}
\newcommand{\ci}{[\ion{C}{1}]}
\newcommand{\stwo}{[\ion{S}{2}]}
\newcommand{\sthree}{[\ion{S}{3}]}

\newcommand{\deepimage}{deep [\ion{Fe}{2}]+[\ion{Si}{1}] image}

\newcommand{\bunit}{erg~cm$^{-2}$~s$^{-1}$~pixel$^{-1}$}
\def\fluxfeii{F_{\rm [Fe\,II]\,1.644}}
\def\fluxsii{F_{\rm [Si\,I]\,1.645}}
\def\fluxhubble{F_{\rm F098M}}
\def\lumfeii{L_{\rm [Fe\,II]\,1.644}}
\def\mfe{M({\rm Fe})}
\def\ffep{f_{\rm Fe}^+}
\def\flevel{f_{\rm a ^4D_{7/2}}} 
\def\fobs{$F_{\rm obs}$}
\def\fextcorr{$F_{\rm ext-corr}$}
\def\stwofifty{$S_{250~\mu {\rm m}}$}
\def\nhx{$N_{\rm H,\,  X-ray}$}
\def\mum{$\mu$m}
\def\vrad{v_{\rm rad}}

\def\halpha{H$\alpha$}

\received{February 15, 2018}
\revised{March 15, 2018}
\accepted{\today}
\submitjournal{ApJ}

\shorttitle{Deep NIR \feii+\sii\ Emission Line Image of Cassiopeia A}
\shortauthors{Koo et al.}

\begin{document}


\title{A Deep Near-Infrared \feii+\sii\ Emission Line Image of the Supernova Remnant Cassiopeia A}

\correspondingauthor{Bon-Chul Koo}
\email{koo@astro.snu.ac.kr}

\author{Bon-Chul Koo}
\affiliation{Department of Physics and Astronomy, Seoul National University \\
Seoul 151-747, Korea}

\author{Hyun-Jeong Kim}
\affiliation{Department of Physics and Astronomy, Seoul National University \\
Seoul 151-747, Korea}

\author{Yong-Hyun Lee}
\affiliation{Department of Physics and Astronomy, Seoul National University \\
Seoul 151-747, Korea}

\author{John C. Raymond}
\affiliation{Harvard-Smithsonian Center for Astrophysics,\\
60 Garden Street, Cambridge, MA 02138, USA}

\author{Jae-Joon Lee}
\affiliation{Korea Astronomy and Space Science Institute, \\
Daejeon 305-348, Korea}
 
\author{Sung-Chul Yoon}
\affiliation{Department of Physics and Astronomy, Seoul National University \\
Seoul 151-747, Korea}

\author{Dae-Sik Moon}
\affiliation{Department of Astronomy and Astrophysics, University of Toronto,\\
Toronto, ON M5S 3H4, Canada}


\begin{abstract}
 
We present a long-exposure ($\sim$10 hr), narrow-band image of the supernova (SN) 
remnant Cassiopeia A (Cas A) centered at 1.644 $\mu$m emission. 
The passband contains \feii\ 1.644 $\mu$m and \sii\ 1.645 $\mu$m lines, and our  
`deep \feii+\sii\ image' provides an unprecedented panoramic view of Cas A, 
showing both shocked and unshocked SN ejecta together with 
shocked circumstellar medium at subarcsec ($\sim 0\farcs7$ or 0.012 pc) resolution. 
The diffuse emission from the unshocked SN ejecta has a form of clumps, filaments, 
and arcs, and their spatial distribution correlates well with that of 
the \spitzer\ \siii\ infrared emission, suggesting that 
the emission is likely due to \sii\ not \feii\ as in shocked material. 
The structure of the optically-invisible western area of Cas A    
is clearly seen for the first time. The area is 
filled with many Quasi-Stationary Flocculi (QSFs) and fragments of the 
disrupted ejecta shell.    
We identified 309 knots in the \deepimage\ and classified them 
into QSFs and fast-moving knots (FMKs).  
The comparison with previous optical plates 
indicates that the lifetime of most QSFs is $\simgt 60$~yr.
The total H+He mass of QSFs is $\approx 0.23 M_\odot$,
implying that  the mass fraction of dense clumps in the 
progenitor's mass ejection immediately prior to the SN explosion is about 4--6\%. 
FMKs in the \deepimage\ mostly correspond to S-rich ejecta 
knots in optical studies, while those 
outside the southeastern disrupted ejecta shell appear Fe-rich.
The mass of the \feii\ line-emitting, shocked dense Fe ejecta is $\sim 3\times 10^{-5} M_\odot$.

\end{abstract}

\

\section{Introduction} \label{sec:intro}

Core-collapse supernova (SN) explosions are  
fundamental phenomena in astrophysics.  
There has been considerable progress in both observational and theoretical studies of  
this phenomenon \citep[e.g.,][and references therein]{jan12,mul17}, 
but our understanding of it, especially the explosion process, is still limited. 

A practical approach to address the problem is to study 
young Galactic supernova remnants (SNRs) where we can  see the footprints of SN explosions. 
Cassiopeia A (Cas A) is one such SNR. It is young \citep[$\sim 340$ yr;][]{tho01,fes06}  and nearby 
\citep[3.4 kpc;][]{ree95,ala14}. 
Its SN type is Type IIb, and the mass of the progenitor star has been estimated to be 
15--25~\msun\  \citep{young06}. 
Our understanding of the explosion of the Cas A SN  
is largely based on the observational studies of {\em shocked} 
ejecta in optical, infrared, and X-ray bands
\citep[e.g.,][and references therein]{fesen01a, hwa04, smi09, del10, ise10, fes11, hwa12, ise12, mil13, mil15}. 
According to these studies, the shocked SN ejecta 
can be characterized into three distinct components:  
the $200\arcsec$-diameter main ejecta ring composed of dense 
($\sim 10^4$~cm$^{-3}$) Ne- and O-burning elements,
the O/S-rich `jet'-`counter-jet' structure extending far beyond the main ejecta ring along 
the northeast-southwest direction, 
and the diffuse X-ray-emitting Fe-rich ejecta plumes stretching beyond the main ejecta ring in the southeast, north, and west. 
The estimated mass of Fe-rich ejecta is 
$\simlt 0.1$~\msun\ which is comparable to the expected 
Fe ejecta mass from complete Si burning \citep{hwa12}.     
Models for three-dimensional  
spatial distribution of these ejecta materials and their spatial relations 
have been developed \citep{del10,mil13}. 

In addition to the shocked ejecta,  
{\em unshocked} SN ejecta in the interior preserving the pristine information
at the time of explosion have been detected.
Alpha elements, e.g., O, Ne, Si, S, and Ar, have been detected in faint mid-infrared (MIR) emission 
by {\it Spitzer} Space Telescope and also in the [S III] $\lambda\lambda$9069, 9531 lines \citep{smi09,del10,ise10,mil15},
while $^{44}$Ti tracing $^{56}$Ni (and therefore its stable daughter nucleus $^{56}$Fe) has been detected 
in hard X-rays \citep{gre14}.
The unshocked S ejecta have a sheet- or bubble-like morphology
suggesting bubbles of radioactive $^{56}$Ni-rich ejecta.
On the other hand, the distribution of $^{44}$Ti is elongated along the east-west direction, 
indicating an asymmetric SN explosion. 
Three dimensional numerical simulations 
demonstrating that  neutrino-driven explosions can explain the shocked diffuse Fe plumes 
as well as the asymmetric $^{44}$Ti distribution have been published 
\citep{orl16,won17}. 

The morphology and nature of the unshocked SN ejecta in Cas A, however, are still uncertain.
Firstly, Fe associated with $^{44}$Ti has not been detected. This could be 
because the associated Fe is diffuse and cool \citep[$\sim 40$ K;][]{del14,leeyh15}, 
but that needs to be confirmed.
Secondly, the  {\it Spitzer} MIR images of O, Si, and S emission lines suggest that 
the unshocked SN ejecta have complex structures that might be directly related 
to the SN explosion dynamics, but their coarse spatial resolution ($6\arcsec$--$10\arcsec$) 
prevents a detailed view of the morphology. 
Low-frequency free-free absorption measurements 
yielded a mass as large as $\sim 3$~\msun\ in the unshocked ejecta  
(\citealt{arias18}; see also \citealt{del14}).   
Finally, there is another form of  
shocked dense ($\sim 10^4$~cm$^{-3}$) Fe 
recently detected in the main ejecta ring with near-infrared (NIR) spectroscopy \citep{leeyh17}. 
Those spectra show essentially only Fe lines, in contrast to 
the spectra of Ne- and O-burning ejecta dominated by S lines, suggesting the presence of 
unshocked Fe ejecta in the interior.  
\cite{leeyh17} further pointed out that the SW shell is prominent in 
NIR \feii\ emission \citep[see also][]{rho03}
and proposed that those regions are likely to be dominated by shocked dense 
Fe ejecta based on comparison between the distribution of the NIR 
Fe line emission and those of the $^{44}$Ti and X-ray Fe line emission.  

Cas A also provides a unique opportunity to study the mass-loss history 
of the progenitor system immediately prior to the explosion.
From the early optical studies of the source, it was
realized that the knots in 
Cas A are composed of two distinct components with very different kinematic and chemical 
properties, i.e., ``fast-moving knots'' (FMKs) and 
``quasi-stationary flocculi''  \citep[QSFs;][]{min59,van70,pei71,van71,kam76}. 
Fast-moving ($\simgt 5,000$~\kms) FMKs are O/S-rich SN ejecta material, while
QSFs, which are almost `stationary' ($\simlt 500$~\kms) 
and bright in H$\alpha$ and \nii$\lambda\lambda$6548, 6583 
emission line images \citep{van71,van85,law95,fesen01b,ala14}, 
are probably dense CNO-processed circumstellar medium (CSM) that has been swept-up by the SN blast wave. 
There are about 40 QSFs scattered both inside and outside the main ejecta shell,
and, from the proper motion studies of QSFs, 
an expansion timescale of $\sim 11,000$~yr was derived \citep{van85}.
QSFs, however, have been studied mainly in the optical band where the extinction is 
very large, i.e, $A_V\simgt 6$ mag \citep[e.g., see][and references therein]{koo17}, 
so that it is quite possible that many other  
QSFs remain undetected. 

In this paper, we present a long-exposure ($\sim 10$ hr) 
image of Cas A obtained by using a narrow-band filter centered at 1.644 $\mu$m line 
with the United Kingdom Infrared Telescope (UKIRT) 3.8-m telescope (hereafter `deep-\feii+\sii\ image').  
In the band, there are two strong lines; \feii\ at 
1.64355271 $\mu$m (a$^4D_{7/2}$ $\rightarrow$ a$^4F_{9/2}$)  
and \sii\ 1.6454533 $\mu$m ($^1D_2$ $\rightarrow$ $^3P_2$ ).\footnote{There is also a H Brackett line $n=12\rightarrow 4$ (1.640688 \mum) in the band. But  it is much weaker than 
\feii\ 1.644 \mum\ line in shocked atomic gas and also most of its contribution will be removed by 
subtracting $H$-band image.}
The \feii\ 1.644 $\mu$m line is one of the strongest NIR emission lines  in
shocked dense atomic gas \citep[e.g.,][]{koo16} tracing 
both shocked SN ejecta material and shocked CSM. 
The \sii\ 1.645 $\mu$m line could be `bright' in unshocked SN ejecta as we will see in this paper.
The organization of the paper is as follows. In \S~\ref{sec:obs}, 
we describe our observations and image processing. In \S~\ref{sec:newresult},
we first briefly discuss the two emission lines in the band.  
Then we present our \deepimage\ and discuss new features 
that have not been seen either in previous low-sensitivity 
images \citep[e.g.,][]{rho03} or in other wave-band images.
In \S~\ref{sec:knot}, we identify QSFs and FMKs in the \deepimage\ and 
present their catalog.  
In Sections \ref{sec:qsf} and \ref{sec:fmk}, we investigate the physical properties of QSFs and FMKs, respectively, 
and discuss the implications. 
Finally, in \S~\ref{sec:conclusion}, we conclude and summarize our paper. 

\section{Observation and Image Processing}  \label{sec:obs}

The narrow-band \feii+\sii\ filter images as well as $H$-band images that we present in this paper   
were taken with the Wide-Field Camera (WFCAM) on the UKIRT 3.8 m telescope 
in 2013 September. 
WFCAM is quipped with four Rockwell Hawaii-II HgCdTe infrared focal plane arrays. 
Each array has $2048\times 2048$ pixels, providing a 
$13.'65 \times 13.'65$ field-of-view (FoV) with a pixel scale of $0\farcs4$/pixel. 
The arrays are arranged in a square pattern with a gap of $12.'83$ between them. 
Since the FoV provided by a single array is significantly larger than the entire Cas A, 
we took images of Cas A flipping between two arrays on the east (arrays 2 and 3) 
so that the exposure while the Cas A is placed on the other array 
can be used for flat-fielding and sky subtraction. 
For each pointing, images were obtained at five jitter positions 
offset by ($\pm 6\farcs4$, $\pm 6\farcs4$)  in RA and Decl.
We performed a $2\times 2$ microstep about each jitter position, 
with an offset size of $1\farcs39$, in order to fully sample the point spread function. 
Five jitter positions with $2\times 2$ micro-stepping give a total of 20 images. 
In summary, a single set of exposures  
provides two sky-subtracted images (one for array 2 and the other for array 3)
with a total per-pixel integration time of $20t$ seconds, where 
$t$ is the exposure time of each image. 
For the \feii\ and $H$ filters, $t=20$ s and 5 s, respectively. 
In the period of 2013 September 4--13, 
a total of 44 sets are obtained for \feii\ and 14 sets for $H$. 
In our observations, we used the same \feii\ filters developed for the UWIFE survey 
\citep{leejj14a}. They have a central wavelength of 
1.642--1.645 $\mu$m with a peak transmittance of $\sim 85$\% 
and effective bandwidth of 0.028 $\mu$m. 

Combining all the images obtained with the two arrays 
results in an \feii+\sii\ image with an effective net exposure time of 10 hrs 
and an $H$ image with 0.8 hrs. This `deep \feii+\sii\ image' has a pixel size $0\farcs2$ and 
its median PSF is $0\farcs7$. 
An astrometric calibration is done by comparing the positions of 
more than 100 bright, isolated stars 
in the field with the 2MASS $H$-band catalog \citep{skr06}.
The $1\sigma$ uncertainty in astrometry is $0\farcs104$ (or 0.52 pixel).
For the photometric calibration, we 
used $H$-band magnitudes of 2MASS catalog stars assuming  
the zero-magnitude flux of $3.27 \times 10^{-8}$ erg cm$^{-2}$ s$^{-1}$ 
for the \feii\ narrowband filter (Lee, Y.-H. et al. in preparation). 
The flux calibration uncertainty ($1\sigma$) is 4\%.
The continuum-subtracted image presented in this paper is produced following 
the procedure in \cite{leejj14a}. 
The sensitivity ($1\sigma$) of the continuum-subtracted deep-\feii+\sii\ image 
is about $2.6\times 10^{-18}$ erg cm$^{-2}$ s$^{-1}$ pixel$^{-1}$. 
The flux of \feii\ emission features derived from the continuum-subtracted image 
will be underestimated because  
there are several bright \feii\ lines in $H$ band (1.5--1.8 \mum; see Table 1 of \citealt{koo16}).
The scaling factor depends on electron density ($n_e$) and temperature 
($T_e$) of the source, e.g., 1.16--1.32 for gas at $n_e=10^3$--$10^5$~cm$^{-3}$ and 
$T_e=7000$~K, and we will be using 1.2 in this paper.

A caveat for our image is that the \feii+\sii\ filter 
has a width of $\pm 2600$~\kms, 
so the high-velocity ejecta could have been missed in our image.  
For example, the bright ejecta filaments in the northern interior have  
high positive and negative velocities \citep{mil13},
and they appear as negative features in our \feii+\sii\ image because we subtracted 
the $H$-band image. But such instrumental `artifacts' are limited to the 
northern central region, and we are seeing most of the ejecta material.

\section{Deep \feii+\sii\ Image and New Features}  \label{sec:newresult}

\subsection{Prelude: Is the Emission \feii\ or \sii?}\label{subsec:prelude}

Before we look at the \deepimage, it would be helpful to have some discussion 
about the two emission lines in the band.
Nebular emission at 1.64 $\mu$m is generally assumed to be the \feii\  
1.644 $\mu$m line. NIR spectra of a wide variety of shock waves in HH Objects and SNRs  
confirm this by showing a wealth of other \feii\ lines, the 
strongest being the one at 1.257 $\mu$m \citep[e.g.,][]{ger01, koo13, leeyh17}. 
However, NIR spectra of SN1987A 
\citep{kjaer10} and the nebular phase of the SN iTPF15eqv 
\citep{mil17} show that the line is mainly
the \sii\ 1.645 $\mu$m line, because the other \feii\ lines are weak or 
absent and also because 
the \sii\ 2$p^2$ $^1D_2$ $\rightarrow$ 2$p^2$ $^3P_1$ companion line at 1.607 $\mu$m  
is present. 
The \sii\ 1.607 $\mu$m line originates from the same upper level as the 
\sii\ 1.645 $\mu$m line so that its intensity ratio to the 
\sii\ 1.645 $\mu$m line is given by the ratio of their Einstein $A$ 
coefficients, which is $(7.14\times 10^{-4})/(2.01\times10^{-3})\approx 1/3$. 

For Cas A, NIR spectra of the 
main ejecta shell and several FMKs and QSFs have been obtained \citep{ger01,koo13,leeyh17}. Their spectra 
show many other \feii\ lines including the strongest 1.257 $\mu$m line but no \sii\ 1.607 $\mu$m line  
except for a few FMKs at a level of $\simlt 10$~\% of the \feii\ 1.644 $\mu$m line intensity 
\citep[see Table 3 of][]{leeyh17}. 
The non-detection of \sii\ 1.607 $\mu$m in most regions suggests that 
the emission from the shocked material in the 
\deepimage\ is dominated by \feii\ emission.
In Appendix~\ref{sec:appendix}, we present the results of shock model calculations for   
the \sii\ line brightness in radiative atomic shocks.  
Note that the \sii\ line is not included 
in most of the shock models in the literature, 
and there is little recent atomic data for the line.
We have scaled 
the \ci\ 9850 \AA\ line flux to estimate the \sii\ brightness,   
which is similar to the \ion{Si}{1} line in the behavior of its excitation rate 
with temperature and in the ionization fraction of \ion{C}{1}. 
We have run shock models with QSF parameters and found  
the flux ratio $\fluxsii/\fluxfeii\simlt 2$~\% in these models. Therefore, for QSFs, the 1.64 $\mu$m 
emission should be dominated by the \feii\ line. 
We have also run shock models for FMKs and found  
$\fluxsii/\fluxfeii\simlt 10$~\% when the abundances of 
Si and Fe are comparable (see Figure~\ref{fig:fig1}1). 
The abundances of Si and Fe in FMKs are not observationally constrained and 
might vary substantially. 
Indeed, among the 58 ejecta knots identified along the main ejecta shell by \cite{leeyh17}, 
only three of them show the \sii\ 1.607 $\mu$m line. 
The inferred $\fluxsii/\fluxfeii$ ratio is 0.9 in one of them while it is  
$\simlt 0.3$ ($3 \sigma$) in the other two. 
Therefore, the Si abundance of FMKs does not seem be significantly greater than their Fe abundance,
and the 1.64 $\mu$m emission from the dense shocked ejecta material in the 
\deepimage\ is probably dominated by \feii\ emission too. 

The diffuse emission from the unshocked ejecta is a different case.  The
ejecta are cold and neutral shortly after the explosion, but they are
subsequently ionized and heated by EUV and X-ray photons from the SNR.  
Models of the time-dependent ionization and heating of the ejecta show that
the gas can reach temperatures of $10^4$ K or higher  
\citep{blair89, sutherland95, eriksen09}, though 
dust was not included in the models, and that could significantly reduce the 
temperature. Observations, however, suggest a much lower temperature, e.g., 
$T_e\simlt 100$~K \citep{arias18}.
 As we discuss in \S~\ref{subsec:interior}, we have no theoretical reason to prefer either the
\sii\ or \feii\ identification for the 1.64 $\mu$m line in the diffuse ejecta.
But because the diffuse emission is closely correlated with {\it Spitzer} \siii\ 
emission and because there is little \feii\ {\it Spitzer} emission in that
region \citep{ise10}, we will assume that the diffuse 
emission is \sii.
We will discuss the emission of unshocked ejecta in more detail in \S~\ref{subsec:interior}.

\subsection{Overall Morphology}\label{subsec:morphology}

\begin{figure}[ht]
\begin{center}
\includegraphics[scale=0.3]{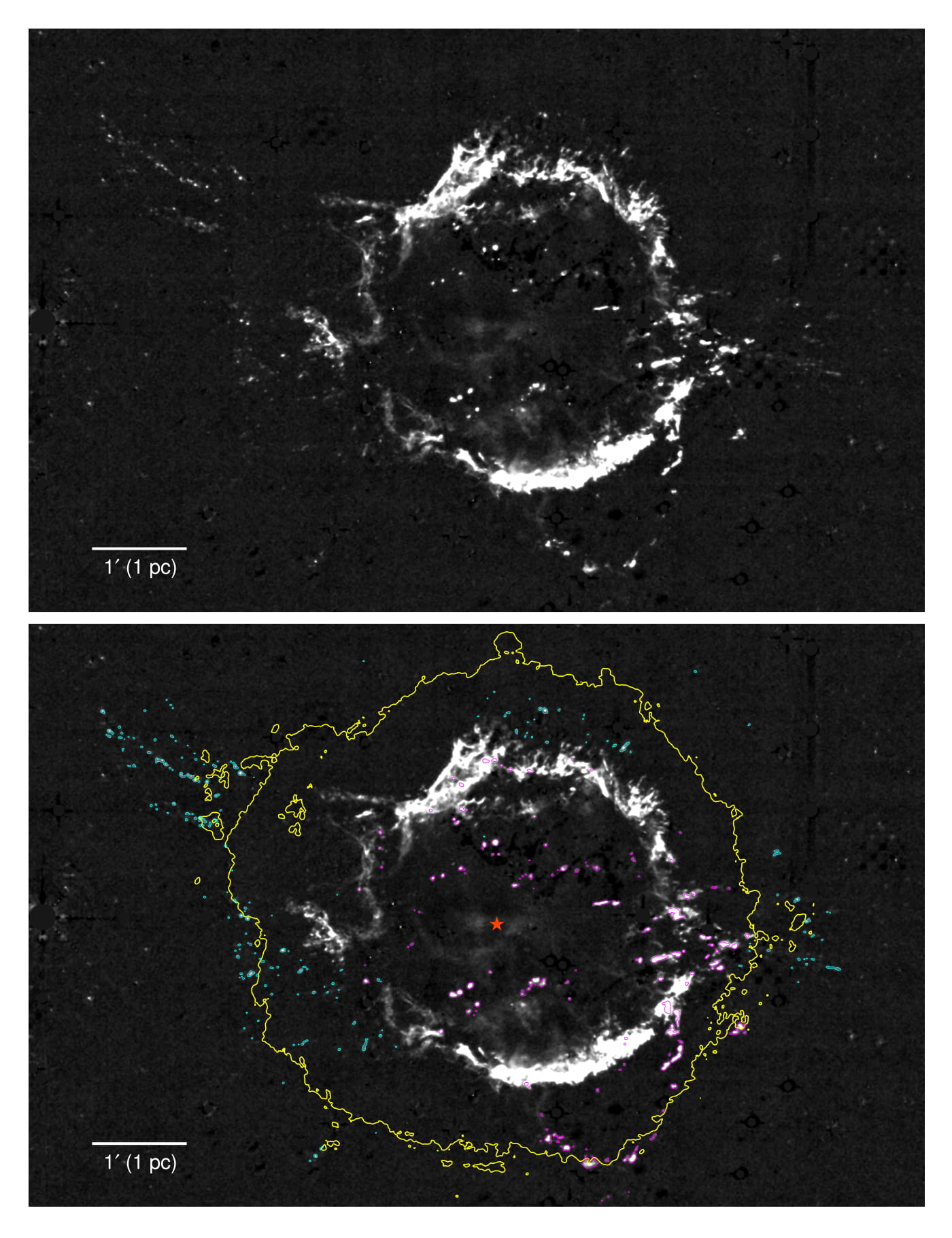}
\caption{
Top: Our deep [\ion{Fe}{2}]+[\ion{Si}{1}] image of Cas A obtained 
from the UKIRT 3.8-m telescope. Stellar sources have been removed. 
The grey scale varies linearly from $-2\times 10^{-18}$ to $2\times 10^{-17}$ 
in erg cm$^{-2}$ s$^{-1}$ pixel$^{-1}$.
Bottom: Same image but with QSFs and FMKs 
marked by magenta and cyan contours, respectively (see \S~\ref{sec:knot}).
The red star represents the explosion center 
at $(\alpha,\delta)_{\rm J2000}=(23^{\rm h} 23^{\rm m} 27.77^{\rm s}, +58^\circ 48' 49.4'')$ 
determined from the proper motion of FMKs \citep{tho01}, while 
the yellow contour marks the outer boundary of the SNR in radio 
corresponding to the intensity level of 0.3 mJy beam$^{-1}$ 
in a VLA 6 cm image \citep{del04}. 
\label{fig:fig1}
}
\end{center}
\end{figure}

Figure~\ref{fig:fig1} shows our deep \feii+\sii\ image of the Cas A SNR.
The overall morphology of the remnant is not very 
different from what has been seen in previous optical or MIR images, 
but, with sub-arcsec angular resolution and continuum subtraction, 
it gives an unprecedentedly clear and panoramic view of the remnant.
In particular, it shows in detail the complex structure of the faint diffuse emission in the
interior that might be from the {\it unshocked} SN ejecta (see also Figure~\ref{fig:fig2}).  
We can see interesting structures, e.g., 
a semicircular arc structure surrounding the explosion center and protrusions from the southwestern ejecta shell  
pointing toward the explosion center, which presumably had been produced during the explosion.   
Figure~\ref{fig:fig1} also reveals the disrupted western part of the main ejecta shell 
that has not been seen in optical observations due to large extinction. 
We can see faint clumps that might be the fragments of a disrupted shell as well as 
bright and faint knots scattered in this area. These new features will be discussed 
in the following sections (\S~\ref{subsec:interior} and \S~\ref{subsec:westarea}).

The most prominent emission feature in Figure~\ref{fig:fig1} is the bright main ejecta shell of $200''$-diameter. 
Its brightness distribution is very similar to those of O, Ne, S, Ar ejecta seen
in optical and \spitzer\ MIR images \citep{ham08,smi09,del10,fes16}, 
so that most of the emission in the main ejecta shell is probably from dense SN 
ejecta swept up by the reverse shock. And  
the emission that we see in the \deepimage\ is probably predominantly \feii\ emission. 
According to one-dimensional SN models \citep[e.g.,][]{woo95}, Fe in the main ejecta shell might be   
composed of $^{56}$Fe atoms of interstellar origin, i.e., $^{56}$Fe atoms 
dumped into the progenitor star during its formation, and $^{58}$Fe atoms synthesized 
by the $s$-process during He core and He shell burning.
In the O-rich core of a 15 \msun\ progenitor, 
the abundances of these two isotopes are comparable 
and $X({\rm O})/X({\rm Fe})\sim 10^3$ in mass \citep[][; see also Figure~\ref{fig:fig5} of Koo et al. 2013]{rau02}.
But three-dimensional simulations suggest that 
it could be mainly Fe newly synthesized in explosive burning and mixed with O-rich ejecta    
due to hydrodynamic chemical mixing \citep[e.g., see Figure~\ref{fig:fig3} of][]{ham10,koo13}. 
The bright southern shell, for example, appears to be dominated by 
Fe-rich ejecta \citep{rho03,leeyh17}. 

Other prominent features in Figure~\ref{fig:fig1} are numerous compact knots scattered 
over the remnant. They are  
QSFs and FMKs.
 The \sii\ 1.645 $\mu$m emission might be negligible
compared to \feii\ 1.644 $\mu$m emission for these shocked 
clumps too, perhaps except some FMKs.
The clumps surrounded by magenta contours in Figure~\ref{fig:fig1} are QSFs (see \S~\ref{sec:knot}).
We can see that the bright clumps are mostly QSFs.   
The most prominent QSF feature is the large 
arc structure in the southwest which was previously detected in  
H$\alpha$ and \nii\ observations \citep{law95,fesen01b,ala14}.
On the other hand, the clustered QSFs
in the western area are newly discovered. 
We discuss the spatial distribution and physical properties of QSFs in \S~\ref{sec:qsf}. 

In Figure~\ref{fig:fig1}, essentially all knots surrounded by cyan contours are FMKs outside the main 
ejecta shell, and 
our continuum-subtracted image provides a clear view of their overall distribution.
The most prominent FMK features are 
the `jet' and `counterjet' structures in the northeastern and southwestern areas, respectively.
These jet structures are enriched with S and they have been studied in detail 
by using optical and NIR forbidden S lines \citep{fes06,ham08,mil13,fes16}.  
They are almost undecelerated and  
thought to have been ejected from the innermost region at the time of core collapse 
by an explosive, jet-like mechanism \citep{fes16}. 
However, no Fe-rich ejecta knots expected in the jet-induced explosion models 
have been reported in the jet area \citep{fes16}.
Our image shows that some FMKs  
outside the disrupted eastern shell appear very bright in \feii\ emission 
while they are relatively faint in the {\it Hubble Space Telescope} 
F098M image obtained by WFC3/IR \citep[][; hereafter the `HST F098M' image]{fes16}. 
The HST F098M image is dominated by ionized S lines, i.e., 
[S III] $\lambda\lambda$9069, 9531, and [S II] $\lambda\lambda$10287--10370, 
so that their difference in appearance 
suggests that these knots could be Fe-rich.
It is worthwhile to note that this is where 
an Fe-rich diffuse X-ray ejecta plume is present \citep{hwa04,hwa12}. 
In \S~\ref{subsec:fmkhst}, we will compare the \feii\ fluxes of FMKs to their HST F098M fluxes. 

\subsection{Diffuse Emission in the Interior}\label{subsec:interior}

\begin{figure}[ht]
\begin{center}
\includegraphics[scale=0.35]{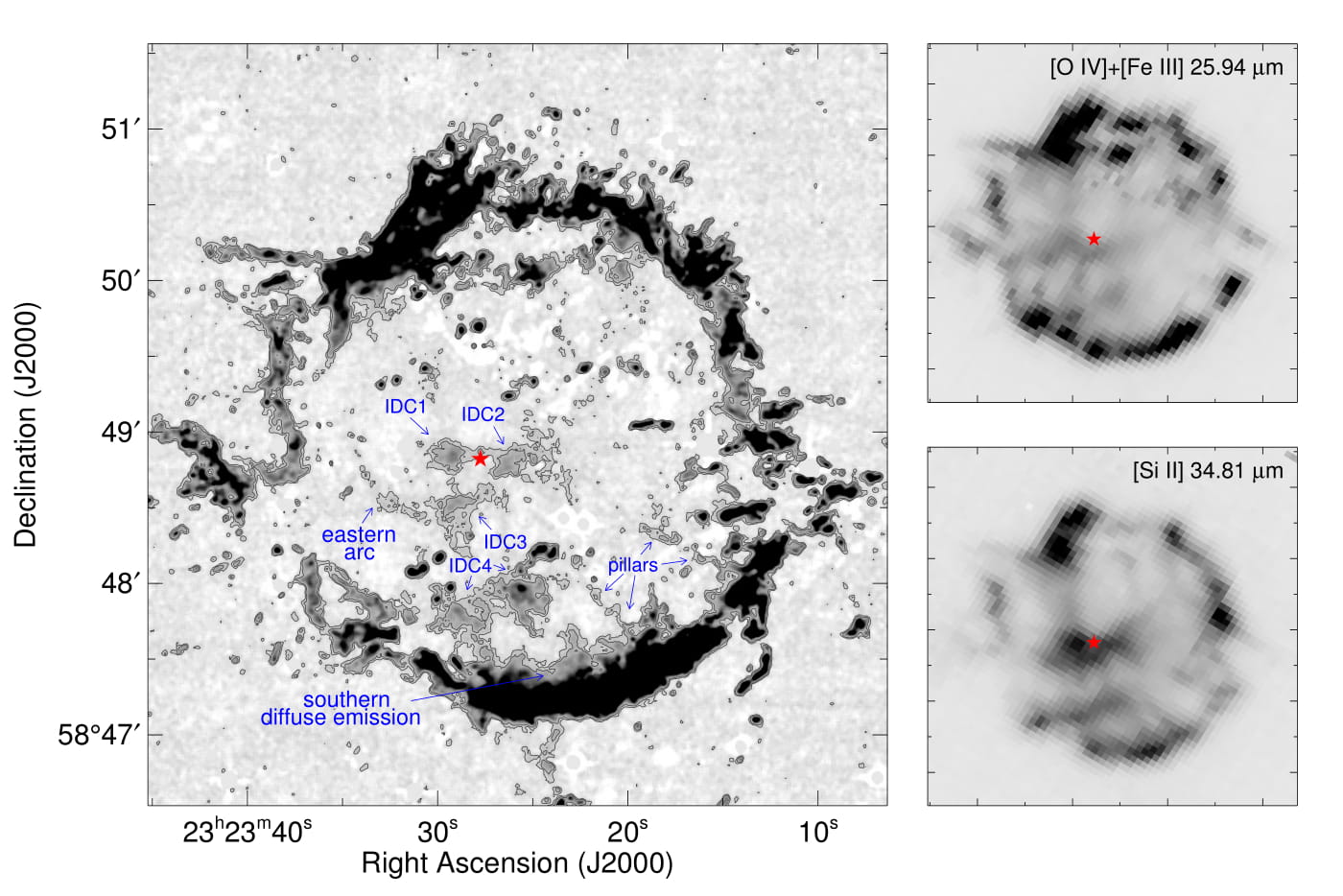}
\caption{Left: Interior diffuse emission of Cas A in 
the \deepimage. Some prominent features are labeled. The red star represents the 
explosion center. The contours are at levels 
$1\times 10^{-18}$ and $2\times 10^{-18}$~\bunit.
Top right: \spitzer\ \oiv+\feii\ 25.9 \mum\ image of Cas A. 
Bottom right: \spitzer\ \siii\ 34.81 \mum\ image of Cas A. 
The \siii\ line is broad ($-4000$ to $+3500$~\kms)
with blue- and red-shifted velocity peaks at $-2,000$~\kms\ and +2,500~\kms\ \citep{ise10}.
The above \spitzer\ images are obtained by integrating over the limited 
velocity ranges to match the velocity range of the \deepimage.
\label{fig:fig2}
}
\end{center}
\end{figure}

Our \deepimage\ reveals for the first time the detailed structure of the 
faint diffuse emission in the interior, well inside the reverse shock (Figure~\ref{fig:fig2}). 
The emission features have the form of clumps, filaments, and arcs, most of which are distributed in the central eastern and southern areas.
Some prominent features are 
(1) interior diffuse clumps (IDC 1 to IDC 4), (2) the semicircular `Eastern Arc'  
of radius $30\arcsec$ surrounding IDC 1,  
(3) `pillars' protruding from the southwestern main ejecta shell, and 
(4) the diffuse emission permeating through the southern main ejecta shell.
The spatial location and morphology  
suggest that the interior diffuse emission is probably from the unshocked SN ejecta, and 
the above features should be closely related to the explosion dynamics.
For example, the location and morphology of the Eastern Arc match 
the bubble-like structure seen in \sthree\ lines \citep{mil15} and 
could be a footprint of a Ni bubble. 
The pillars are also interesting.
There are at least two pillars, each of which is  $30''$--$45''$ (or 0.49--0.74 pc) long, and they are 
pointing almost exactly to the explosion center. 
The morphology of these pillars is similar to the Rayleigh-Taylor fingers developing during the explosion at the Si/O interface, although they appear to point inwards not outwards as shown in numerical simulations \citep{kif03}.  
 
If  the 1.64 \mum\ emission is \feii\ 1.644 \mum\ emission 
from Fe synthesized in SN explosion, we may expect to see some correlation 
with the $^{44}$Ti emission, because they are both products of $\alpha$-rich Si burning. 
The distribution of 1.64 \mum\ emission, however, shows no correlation with the $^{44}$Ti emission 
which is distributed along the east-west direction \citep{gre14}.
Instead, the overall spatial distribution of the 1.64 \mum\ emission matches well with that of the faint 
\siii\ 34.81 $\mu$m emission seen in \spitzer\ observations (Figure~\ref{fig:fig2}), 
which was proposed to be from unshocked SN ejecta 
heated by UV/X-ray emission from the reverse shock \citep{ise10}.
Furthermore, \feii\ 17.94 $\mu$m and 25.99 $\mu$m lines were not detected   
in the interior in \spitzer\ observations  
\citep[][see also Figure~\ref{fig:fig2}]{smi09,ise10,del10}. 
There is strong  \oiv +\feii\ 25.9 \mum\ emission, but \cite{ise10} showed that 
its line profile matches well with the other line profiles  
when we assume that the emission is entirely from \ion{O}{4}. 
These MIR characteristics suggest that 
the interior diffuse emission is mostly, if not all, 
\sii\ 1.645 $\mu$m emission, not \feii\ 1.644 $\mu$m emission. 

On the other hand,  Si atoms (I.P.=8.15 eV) in the unshocked SN ejecta 
are expected to be ionized by UV photons from the reverse shock, which will be dominated by  
\ion{O}{6} doublet at $\lambda\lambda$1032, 1037 \citep{hamilton88}. 
These UV photons may be attenuated by SN dust 
in the Cas A interior, so the inner part of the ejecta can be shielded. 
NIR to submillimeter observations indicate that there is a large amount ($\ge 0.1$--0.2~\msun) of 
newly-formed dust mixed with unshocked SN ejecta in the interior 
\citep{dun09,sib10,bar10,are14,delooze17}. 
Si atoms can be also 
ionized by X-ray photons, but if there are not enough photons, 
some fraction could be in a neutral form. 
According to low-frequency radio observations, the characteristic temperature of the  
unshocked SN ejecta is $\simlt 100$ K \citep{arias18}, in which case
the collisional excitation of \sii\ 1.645 \mum\ line is very inefficient 
because of the Bolzmann factor $\exp(-8760/T)$, 
and the emission from recombination 
would probably dominate. Since neither the abundances nor 
the ionization states of elements in the diffuse
emission region is known, we have no theoretical reason to prefer 
either the \sii\ or \feii\ identification for the 1.64 $\mu$m line in the diffuse ejecta.
However, based on the fact that diffuse emission is strongly 
correlated with the {\it Spitzer} 
\siii\ emission and that there is little {\it Spitzer} \feii\ emission in that
region \citep{ise10}, we tentatively conclude that the diffuse emission is \sii.\footnote{In an accompanying paper \citep{raymond18}, we explore the origin of the interior diffuse emission using NIR spectroscopic data.}   

\subsection{Western Area}\label{subsec:westarea}

\begin{figure}[ht]
\begin{center}
\includegraphics[scale=0.35]{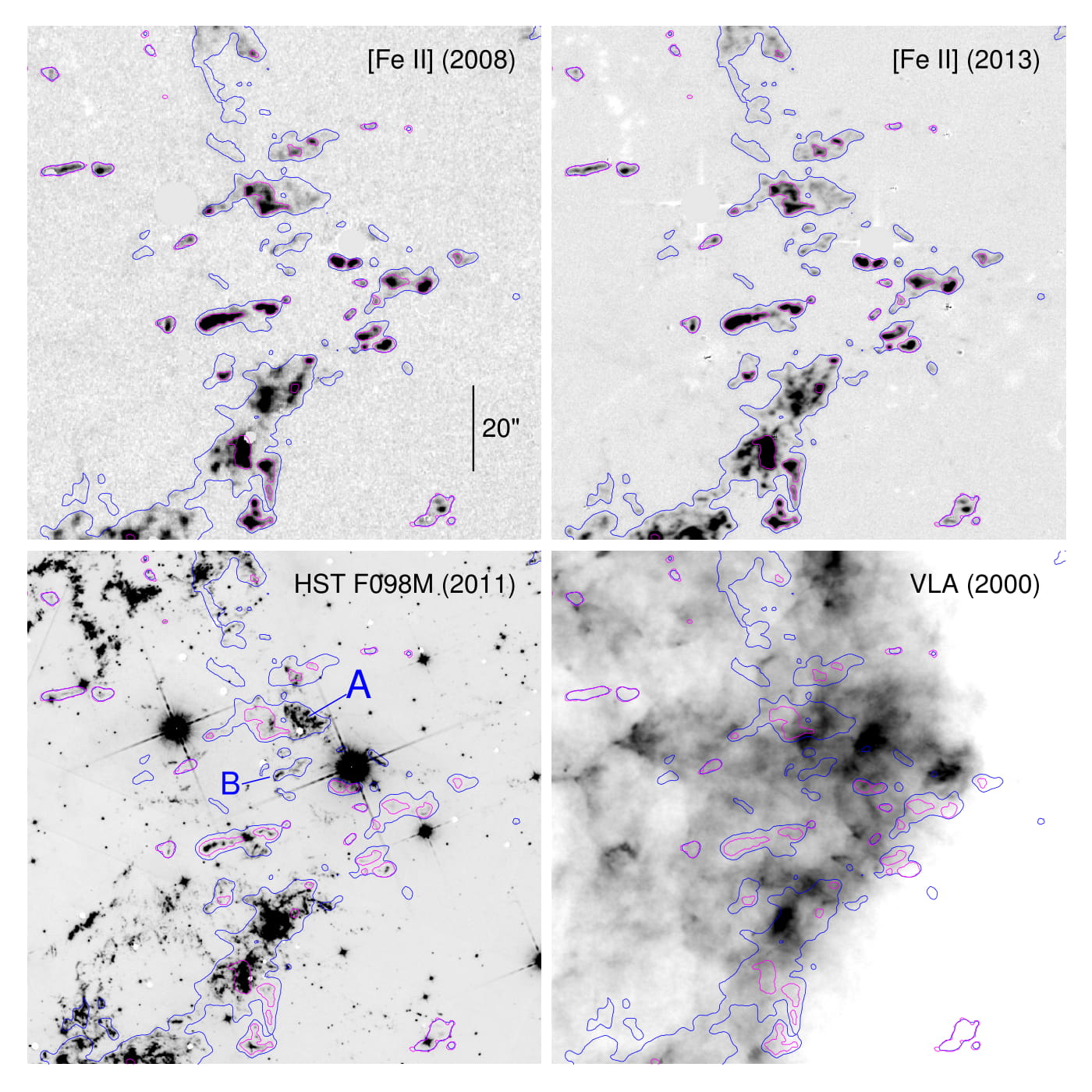}
\caption{Enlarged view of the western area of Cas A:   
[\ion{Fe}{2}] +[\ion{Si}{1}] image obtained in 2008 August \citep{leeyh17},  
\deepimage\ obtained in 2013 September,
HST F098M image obtained in 2011 November \citep{fes16}, and 
VLA 6 cm image \citep{del04}.
The magenta contours mark the boundary of QSFs identified in the \deepimage\ 
(see \S~\ref{sec:knot}), while the blue contours  
represent the areas brighter than $4 \times 10^{-18}$~\bunit\  
in the same image.  
\label{fig:fig3}
}
\end{center}
\end{figure}

Figure~\ref{fig:fig3} gives an enlarged view of the western area, 
showing the complex and clumpy structure of this region.
Over a $1'$-long arc, the bright main ejecta shell is missing but  
filled with clumps of different brightness that are 
$\simlt 20''$ in size, 
some of which are well beyond the main ejecta shell.
The brightest clumps have relatively sharp outer boundaries while 
the faint clumps appear fuzzy. 
Some bright clumps appear embedded in an extended faint envelope.  
In Figure~\ref{fig:fig3}, we show another \feii\ image of Cas A obtained in 2008 
\citep[][see \S~4.2]{leeyh17} for comparison. The 2008 image has a lower sensitivity, 
so that faint clumps in the \deepimage\ are not visible. 
For the clumps seen in both the 2008 and 2013 images, we 
measure their proper motions and explore their nature, i.e., FMK versus QSF, in \S~\ref{sec:knot},
but from Figure 3, it is clear that  
the bright clumps (marked by magenta contours) 
do not show measurable proper motions, indicating that they are QSFs.
The nature of these clumps can be confirmed 
by comparing to the HST F098M image which is also shown in Figure~\ref{fig:fig3}. 
For example, the large clump marked as Clump A is bright in the Hubble image but faint in 
\feii\ as are the small clumps marked as Clump B. 
These clumps are located in the missing portion of 
the main ejecta shell, so that they are likely the 
fragments of the disrupted ejecta shell. 
The overall morphology is indeed similar to that of the fragmented shell 
disrupted by dense ejecta knots launched just outside the Fe core in 
numerical models \citep[e.g., see Figure~\ref{fig:fig10} of][]{orl16}.  
On the other hand, the bright clumps marked by magenta contours 
are only barely visible in the HST F098M image, supporting their 
QSF nature because QSFs are faint in [\ion{S}{2}] and [\ion{S}{3}] lines. 
Some large QSFs appear to be embedded 
in a diffuse faint envelope of ejecta material, which seems to indicate 
their physical interaction.   
In \S~5.2, we discuss the lifetime of QSFs and 
show that most QSFs are likely to survive until they encounter the ejecta shell, 
where they experience a strong shock and destroyed. 
These QSFs, if they are indeed interacting with ejecta material, 
must have been relatively recently swept up by the reverse shock. 

The western area is 
where radio synchrotron emission shows distinct characteristics. 
In this region, the radio brightness is enhanced as we can see 
in the VLA image in Figure~\ref{fig:fig3}.
The spectral index is steep and the motions of radio-bright knots are significantly non-random
with some knots moving inward, indicating interactions of SNR blast wave with dense medium 
\citep{and95,and96,keo96}.
The electron density in this region derived from X-ray analysis, 
however, is not particularly large \citep{hwa12}. 
There is a molecular cloud at $\vlsr=-40$~\kms\ 
superposed to this area, and it has been 
proposed that this molecular cloud is responsible for such 
distinct characteristics \citep{keo96,kil14,kil16}.
Most of the molecular gas, however, lies in the foreground of Cas A
\citep[see][]{kra04,wil05,dun09}.  
Also we do not see \feii\ emission due to the blast wave-molecular cloud interaction  
as we often see in other SNRs interacting with molecular clouds. 
(Nor do we see H$_2$ 2.122 $\mu$m emission in our 600-s integration H$_2$ narrow-band image obtained by 
using the Wide-field Infrared Camera attached to the Palomar 5 m telescope.)
The large population of QSFs in our \deepimage\ instead suggests a  
denser CSM there. Hence, another possible 
explanation for the anomalous radio properties in the western area seems to be  
the interaction with dense clumpy CSM and the shell fragments.  
It is well established that, if the medium is clumpy, the flow behind a strong shock becomes turbulent and magnetic field can be amplified \citep[e.g.,][]{ino09}, so we expect to see enhanced radio synchrotron emission. 

\section{Identification and Classification of Knots in the Deep [Fe II]+[Si I] Image}\label{sec:knot}
\subsection{Identification of Knots}

We identified knot features in the \deepimage\  in an automated manner.
We first smoothed the continuum-subtracted, \deepimage\ using SMOOTH function
in IDL that makes a smoothed image with a boxcar average of the specific 
width, three pixels in this case, and estimated the background of the image 
to determine a threshold for the identification of knots. 
The background value and standard deviation ($\sigma$) estimated from 
several source-free regions of the image is  
$(-0.35 \pm 8.0) \times 10^{-19}$~\bunit.
We detected the knot features outside (i.e., not-superposed on) the main ejecta shell
using the threshold of $4 \times 10^{-18}$~\bunit\ (or 5$\sigma$ above the background), 
a threshold that is low 
enough to include faint emission but high enough to exclude background noise.
We have drawn the contours on the image by the threshold and defined
them as ``knots'' after excluding residual features left from 
continuum subtraction, artificial patterns from the detectors, and 
the small contours with less than five pixels considering the pixel scale 
($0.2\arcsec$) and seeing ($\lesssim 1\arcsec$).
The total number of knots detected and catalogued outside the main ejecta shell is 258.

We fitted each contour by an ellipse 
using the IDL procedure FIT\_ELLIPSE 
from Coyote IDL Library\footnote{The {\it Fit\_Ellipse} is a part of
the Coyote IDL Program Libraries (\href{http://www.idlcoyote.com}
{http://www.idlcoyote.com}).} 
to derive their geometrical parameters, i.e.,  
the central coordinates, major and minor axes, 
and position angles (P.A.; from north to east), 
while the area and \feii\ flux were directly   
estimated from the contours. 
The uncertainty in flux estimation is $\sim 10\%$. 
The derived parameters are listed in Table~\ref{tbl-1}.

\subsection{Classification of Knots}\label{subsec:knotclass}

The identified knots have been classified by finding their counterparts in   
another \feii\ image of Cas A obtained in 2008 and by measuring their proper motions. 
The 2008 image was obtained on 2008 August 11 by 
using the Wide-field Infrared Camera attached to the Palomar 5 m telescope \citep{leeyh17}.
The total integration time of this 2008 image is 5400 s and its sensitivity is 
about a factor of 2 lower than that of the \deepimage.
The seeing of the 2008 image 
is also worse with $0\farcs9$ compared to $0\farcs7$ of the 2013 image.  
These restrictions limit the identification to relatively bright knots.
The knots in the 2008 image were identified and cataloged by the same method
applied to the 2013 image.
The threshold for the knot identification was 
$1.14 \times 10^{-17}$~\bunit\ (or 4$\sigma$ above the background)
from the estimated background value and standard deviation 
was $(0.47 \pm 2.7) \times 10^{-18}$~\bunit. 
The total number of identified knots is 92, only about 40\% of
the number of knots in the 2013 image.

\begin{figure}[ht]
\begin{center}
\includegraphics[scale=0.55]{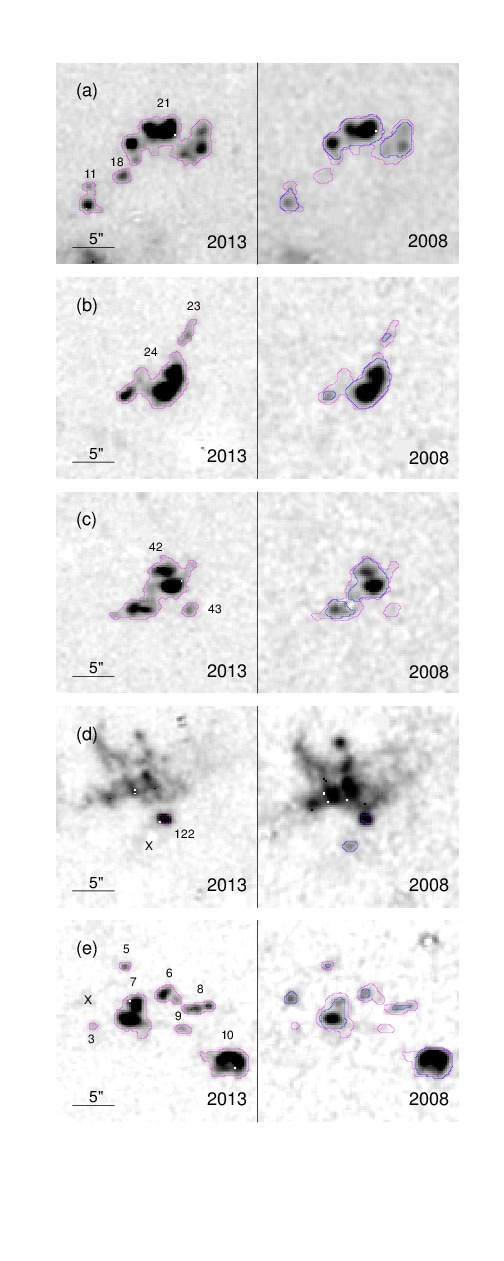}
\vspace*{-25mm}
\caption{Knots that required special care in finding counterparts in the 2008 and 2013 
\feii+\sii\ images (see \S~\ref{subsec:knotclass}). Blue and red contours mark   
the identified knots in the 2008 and 2013 images, respectively. 
(a)--(c) Knots identified as two separate knots in the 2008 image but 
identified as a large single knot in the 2013 image.  
(d)--(e) Knots well defined in the 2008 image but not apparent in the 2013 image. Their expected positions assuming no proper motion are marked by `x' in the 2013 image. In (e), Knot 6 is an example where knot has a faint tail in the 2013 image but is detected as a compact knot in the 2008 image, so that the proper motion is  overestimated in the automatic search. 
\label{fig:fig4}
}
\end{center}
\end{figure}

The counterparts of the knots in the 2013 image 
were searched for by cross-correlating the two catalogs.
Considering that there are knots with little proper motion as well as knots 
with a large proper motion, 
we first have extracted the pairs showing little proper motion between 2008 and 2013, which 
are the knots with the distance between the two intensity-weighted centers 
less than 0.9$\arcsec$ (see below). 
For the latter, we have found the pairs of knots with the smallest distance and
visually inspected to confirm that they are counterpart of each other 
based on their shapes and
relative positions in the 2013 and 2008 images.
In the crowded regions where the 2013 and 2008 knots often overlap along the expansion direction 
(e.g., the NE jet region or the eastern part of the shell), 
we shifted the 2008 knots by $\sim 2\farcs5$ in the radial direction from 
the remnant center and found the counterparts to prevent mis-identification.
All but seven of the 92 knots in the 2008 images have their counterparts
in the 2013 image.
Among the seven knots with no counterpart in 2013,
two are small and faint, so they may not be real emission; 
three are large knots in the 2013 image that are 
detected as a couple of separated knots in the 2008 image because of lower sensitivity 
(Figure~\ref{fig:fig4} (a)--(c)), so
they have been already extracted as 2008 knots matched better with smaller distances; 
the remaining two are rather well defined in the 2008 image, 
so they could be QSFs that had disappeared in 2013 (Figure~\ref{fig:fig4} (d)--(e)).
For the knots with counterparts, 
we estimated the proper motion from the distances between the two central positions and 
the position angles defined as the position in 2013 with respect to the position in 2008
(from north to east) using their central coordinates. 
As some knots are extended or irregular rather than knot-like,
we derived the proper motion using the intensity-weighted centers as well as
using the geometrical centers from the fitted ellipses.

Previous optical studies showed that the proper motions of 
QSFs are mostly $\simlt 0.02$ arcsec yr$^{-1}$ although a few QSFs 
show unusually large ($\simlt 0.18$ arcsec yr$^{-1}$) proper motions \citep{van71,van83,van85}.
On the other hand, FMKs do have large proper motions, e.g., 
$0.31$ arcsec yr$^{-1}$ for an expansion velocity of 5000~\kms\ 
corresponding to a positional shift of 
$\sim1\farcs5$ between the 2008 and 2013 images. 
We classified the knots with little proper motion, i.e., the knots with the distance 
between the intensity-weighted centers in 2013 and 2008 less than $0\farcs9$, as QSFs. 
The $0\farcs9$ of the positional shift corresponds to a proper motion of 0.18 arcsec yr$^{-1}$,
the largest value reported in \citet{van85} and rather large compared to typical proper 
motions of QSFs, but it is appropriate when we consider
the uncertainty of center positions of the detected knots between the 2013 and 2008 images.
Since the 2013 image has higher resolution and sensitivity, 
some emission is only barely detected in 2008, making it difficult to find 
the exact center positions and leading to the overestimation of the positional shifts, so
those QSFs can be missed by a smaller shift criterion.
For example, an elongated emission with a bright knot and a faint tail in the 2013 image
is only detected as a compact knot in the 2008 image, so the positional shift
estimated between the center of an elongated ellipse and the center of a circle-like ellipse
can be overestimated (e.g., Knot 6 in Figure~\ref{fig:fig4} (e)).
The classified QSFs,
although we used a conservative criterion, show the positional shift
less than $0\farcs54$ (or 0.11 arcsec yr$^{-1}$) with the median of $0\farcs084$ (or 0.017 arcsec yr$^{-1}$),
and we visually confirmed that the ones with a relatively large positional shift indeed 
have different shapes between the two images and their proper motions have been 
likely overestimated.
Among the 85 knots with counterparts, 43 knots  
are classified as QSF knots, and the other 42 knots 
with large proper motions are classified as FMKs.
This leaves 173($=258-85$) knots without counterparts. 

We further classified these 173 knots without 2008 counterparts using the HST F098M and 
HST F850LP images. The latter HST image was obtained  
in 2004 using ACS/WFC with the F850LP filter \citep{fes06}. Since FMKs are generally 
seen as bright in these HST images including ionized S lines, we classified the knots 
that have a counterpart in either HST image with an obvious proper motion as 
FMKs. By comparing
with the HST image, 125 knots were classified as FMKs.
The remaining 48 knots without a counterpart in the HST images were classified as 
FMK candidates (12 knots) if they are located around the NE or the western jet area, or 
as QSF candidates (36 knots) if they are located around the region where QSFs are mainly distributed 
or have a counterpart in the H$\alpha$+\nii\ and/or \nii\ image \citep{van85,ala14}.
As the FMK/QSF candidates are generally faint, another deep imaging observation with  
the narrow-band [\ion{Fe}{2}]+[\ion{Si}{1}] filter will be necessary to confirm their origin.
In summary, among 258 knots detected in the 2013 image,
we classified 79 QSFs including 36 candidates and 179 FMKs including 
12 candidates. The classification of the knots is presented in Table~\ref{tbl-1}.

\subsection{Knots superposed on the main ejecta shell}\label{subsec:knotshell}

There are also knot-like features superposed on the main ejecta shell.  
Indeed, the shell is very clumpy, so that, on the shell, defining a ``knot'' is subjective 
and cataloging all those knot-like features may not be meaningful. 
We limit the identification to bright QSF features.
Since the ejecta shell itself is so bright that its boundary was defined by 
the threshold ($4 \times 10^{-18}$~\bunit) used for the knot identification,
we identified brighter knot features embedded in the shell by higher threshold values. 
Note that the `shell' defined by this boundary 
includes some diffuse clumps beyond the main ejecta shell 
in the western area (see Figure~\ref{fig:fig6}), but we may 
consider them to be part of the disrupted and fragmented shell, 
similar to those in the eastern area. 
On the shell, 164 emission features were detected with 
the threshold values from $2.4 \times 10^{-17}$ to $4.0 \times 10^{-16}$~\bunit.
These threshold values are rather arbitrary but have been determined to identify bright, 
knot-like features via visual inspection.
Among the detected emission features, since our purpose here 
is to identify QSFs, we excluded the ones with 
extended and/or filamentary structure that trace the main shell and show 
obvious proper motion between 2013 and 2008 images---the ones with
little proper motion have not been excluded even if they have irregular or 
elongated shapes because they are supposed to be QSFs. 
We also excluded the knots on the shell that are below the applied threshold 
and faint knots close to the shell whose proper motions are similar  
to the proper motion of the main shell.
Finally, we identified 51 QSFs detected in this way and included them 
in the knot catalog with a superscript `s' (Table~\ref{tbl-1}).

\subsection{A Catalog of Knots and Extinction Correction}

Table~\ref{tbl-1} is the final table of 309 knots ($=130~{\rm QSFs}+179~{\rm FMKs}$) 
identified in the \deepimage.
We numbered the knots by their P.A. with respect to the explosion center and
distance from the explosion center. In the case of P.A., we sorted the knots by an interval of 5$\arcdeg$ 
so that the knots at a similar distance have been numbered in a sequence, for visual convenience in a finding chart. In Table~\ref{tbl-1}, the geometrical parameters, i.e.,     
position of the geometrical center, major and minor axis, 
and $\psi_{\rm ellipse}$ (= position angle of the major axis measured from north to east) 
are from ellipse fitting, while   
the area and  the observed  \feii\ 1.644 $\mu$m fluxes \fobs\ are 
directly estimated from their contour boundaries. 
The proper motion parameters ($\mu$ and $\theta_\mu$) are obtained from 
the intensity-weighted central positions in 2008 and 2013. 
In the last column, the classification is given. 

\begin{figure}[ht]
\begin{center}
\includegraphics[scale=0.35]{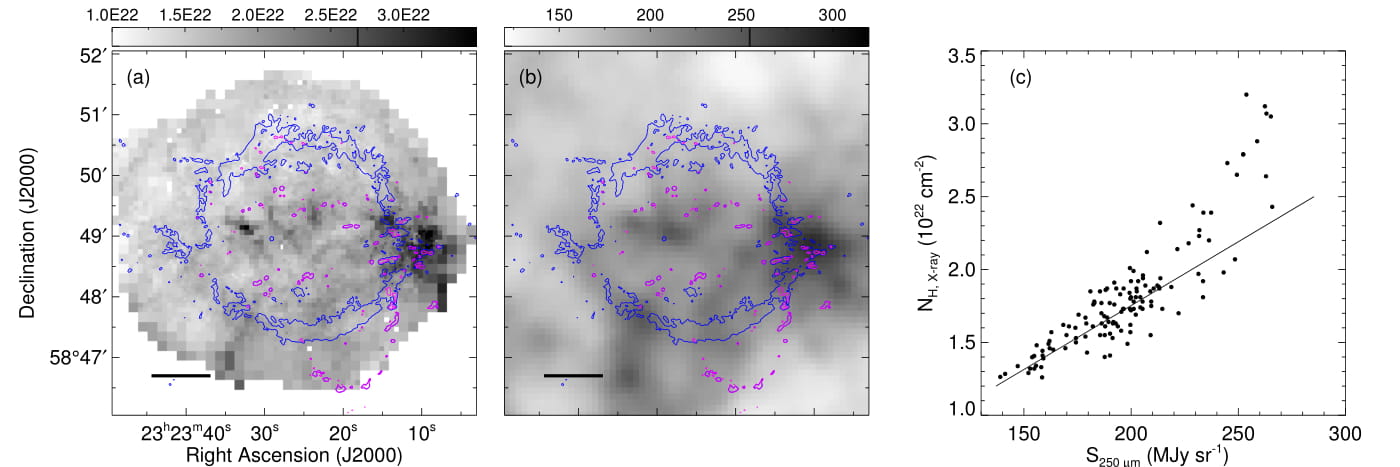}
\caption{(a) Column density map of H nuclei from X-ray  \citep[$N_{\rm H, X-ray}$;][]{hwa12}.  
The scale bar is in units of cm$^{-2}$. The locations of QSFs are marked by magenta contours.
The blue contour represents the main ejecta shell. 
(b) \herschel\ image of 250 $\mu$m brightness ($S_{250~\mu{\rm m}}$). 
The scale bar is in units of MJy sr$^{-1}$.
(c) $N_{\rm H, X-ray}$ versus $S_{250~\mu{\rm m}}$ toward QSFs. 
The solid line shows the relation for dust emission at $T_d=20$~K (see text).
\label{fig:fig5}
}
\end{center}
\end{figure}

Table~\ref{tbl-1} also lists extinction-corrected \feii\ 1.644 $\mu$m fluxes  \fextcorr.
The extinction to Cas A is large 
and varies significantly over the field, e.g., $A_V=$7--15 (see below). 
Column density maps of foreground 
gas/dust across the Cas A SNR have been 
obtained in radio and X-ray \citep{rey02,hwa12}.
We used the column density map of \cite{hwa12} from 
X-ray spectral analysis.
We adopted $A_V/N_{\rm H}=1.87\times 10^{21}$ cm$^{-2}$ mag$^{-1}$ 
and $A(1.644~\mu{\rm m})/A_V=  1/5.4$, which is for the dust opacity of the general ISM \citep{dra03}.
Several QSFs in the southernmost area and FMKs in the NE jet area are 
outside or crossing the boundary of the column density map. For those knots, we 
used the \herschel\ SPIRE 250 $\mu$m image for extinction correction.
The extinction toward Cas A is mostly due to the ISM in the 
Perseus spiral arm, so that the dust emission at 250 $\mu$m is a 
good measure of the extinction to Cas A \citep[e.g.,][]{delooze17}.  
This is confirmed in Figure~\ref{fig:fig5}, where 
we compare  the \herschel\ SPIRE 250 $\mu$m brightness (\stwofifty) to the 
X-ray absorbing column density (\nhx) at the positions of QSFs.
They are well correlated, and 
for $N_{\rm H,\, X-ray}\simlt 2\times 10^{22}$ cm$^{-2}$, 
their relation is linear and is consistent with the 250 $\mu$m emission 
being from 20 K dust with the general ISM dust opacity, i.e., 
$S_{250~\mu {\rm m}}/N_{\rm H,\, X-ray} \approx 1.1~{\rm MJy}/10^{20}~{\rm cm}^{-2}$.   
Towards those QSFs/FMKs outside the X-ray column density map, \stwofifty$\approx 150$~MJy, 
so that we can use this relation.  For higher column densities, \stwofifty\ are considerably smaller than 
the brightness expected from \nhx. This could be because 
the high column densities are due to molecular clouds  
and the temperatures of dust associated with molecular clouds are lower.
Alternatively, it could be because the X-ray absorbing columns are overestimated.
It was pointed out that the extinction from the X-ray analysis appears to be 
somewhat  ($\simlt 0.5$~mag at 1.644~$\mu$m) higher than those from other studies \citep{leeyh15},
in which case the extinction-corrected fluxes of the heavily extincted knots  
in Table~\ref{tbl-1} could have been overestimated by $\simlt 60$\%. 

\section{Quasi-Stationary Flocculi in [\ion{Fe}{2}] Emission}\label{sec:qsf}

\subsection{QSF Finding Chart and Their Optical Counterparts}

Figure~\ref{fig:fig6} is the finding chart of 130 QSFs in Table~\ref{tbl-1}. 
The QSFs with their radial velocities known from  
optical observations by \cite{ala14} are filled with colors based on their radial velocities. 
We have searched for optical counterparts of QSFs by directly 
comparing the \deepimage\ with the \nii\ $\lambda$6583 line 
image of \cite{ala14} via visual inspection instead of cross-correlating the two catalog 
because QSFs in the two images have different, irregular shapes.
For those with counterparts,  we present the IDs and 
radial velocities of \cite{ala14} in Table~\ref{tbl-2}. 
Some QSFs have subknot structures in the \nii\ $\lambda$6583 image; 
in this case, we present the mean velocity of the subknots.
Note that the QSFs in the western area are not covered in the \nii\ $\lambda$6583 image.
Among the 130 QSFs, 39 are found to have optical counterparts.  
In some cases, two QSFs correspond to one \nii\ knot, so that about 80\% of  the 
44 \nii\ knots in the catalog of \cite{ala14} are detected in \feii\ emission. 
The other 20\% of the \nii\ knots are either faint or disappeared.
Several knots clearly appear to have disappeared, 
e.g., A6, A26, A30, and A32 in their catalog.
The \nii\ image was obtained in  
2009 September and 2011 December \citep{ala14}, 
so this fraction appears rather high 
compared to their lifetime ($\simgt 60$ yr, see below).
But considering the non-uniform spatial distribution and the short time interval, 
this may not be statistically significant.


In Table~\ref{tbl-2}, we also present the ID, proper motion, and radial velocity for the QSFs 
detected by \citet[][their Table 2]{van85} in H$\alpha$+\nii\  emission. 
We compared the \deepimage\ with Figure~4 of \citet{van85} 
and Figure~7 of \citet{van83}. These images had been obtained in 1983 and
1976, respectively. (Five QSFs in the southernmost area, i.e., 
R36--R40 in \citealt{van85}, are shown in the latter figure.) 
In the case that a \feii\ knot is matched with several 
optical knots, we present the mean values of proper motion and radial velocity.
Three of the QSFs detected by \cite{van85} were not detected by \cite{ala14}, so that 
the total number of QSFs with optical counterparts is 42. 
The remaining 88$(=130-42)$ QSFs in Table~\ref{tbl-1} are new QSFs identified in this work. 
They are the ones surrounded by empty contours in Figure~\ref{fig:fig6}.
About 30\% of them are located in the disrupted western portion of the main ejecta shell where the  
extinction is large ($A_V\ge 10$ mag), and most of them might have been unseen in previous optical observations 
because of large extinction. The faint ones could have been also undetected previously because of    
low sensitivity. On the other hand, there are bright newly-detected QSFs 
and they might have appeared between 2011 December and 2013 September.  
For example, the bright Knot 7 and the surrounding knots are not seen in either 
\cite{ala14} or \cite{van83} (see \S~5.3). 
It is worthwhile to note that many newly detected knots in the northeastern central region
appear to be lying on an arc structure 
that connects to the prominent QSF arc structure in the south. 

Most of the 42 QSFs in Table~2 seem to have quite long-lived.  
We could not confirm the presence of all 42 QSFs in old plates because of 
their lower sensitivity. Instead we checked if  the 
QSFs identified in previous studies are still visible in our 2013 image. 
\citet{baa54} presented a broad-band red plate taken in 1951 where they 
identified two (\#2 and \#3) bright QSF features.
They correspond to QSF 141/146 and 301/303 in our catalog. 
There are also about dozen faint knot features in the image, and  
only one of them has disappeared. 
\cite{van71} presented a catalog of 19 QSFs marked on 
a broad-band red plate taken in 1968. 
(They are R1--R19 in \citealt{van85}.)
Three of them (R6, R15, and R16) are not visible in our image.
(Two knots, R8 and R18 are superposed on the shell and not clear.) 
\cite{van85} presented a catalog of 40 QSFs marked on 
an 1983 image. Three of them (R15, R31, R32) are seen 
neither in our \deepimage\ nor in the 2009/2011 image of \citet{ala14}. 
One knot (R31/A32) is seen in \cite{ala14} but not in our image. 
\cite{van85} compared deep broad-band red plates (H$\alpha$+\nii) 
taken in 1958 and 1983 and found that only one of 
25 QSFs seen on the 1958 plate disappeared, and they concluded  
that most QSFs have lifetimes $\simgt 25$~yr. 
Our comparison above shows that the lifetimes of QSFs 
are indeed significantly longer than this, e.g., $\simgt 60$~yr (see also \S~5.2).

\subsection{Physical Properties of QSFs}

In Figure~\ref{fig:fig1}, QSFs are marked by magenta contours, and we can see that  
the bright knot features in the \deepimage\ are mostly QSFs. 
This is shown quantitatively in Figure~\ref{fig:fig7} (a), which 
shows the observed \feii\ 1.644 $\mu$m flux distribution of the identified knots.
The filled bars represent QSFs in Table~\ref{tbl-1}, while the empty ones are FMKs. 
The median of the observed fluxes (\fobs) of QSFs is  $1.86\times 10^{-15}$ ergs cm$^{-2}$ s$^{-1}$.
For comparison, the median \fobs\ of the FMKs is $0.24\times 10^{-15}$ ergs cm$^{-2}$ s$^{-1}$.
The QSF knots are also generally large. 
Their geometrical mean radius ranges $0.\arcsec28$--$4.\arcsec39$ with a median of 
$1\farcs03$ (cf. $0\farcs61$ for FMKs).
The large fluxes of the QSFs, however, 
are not entirely due to their large areas as we see in Figure~\ref{fig:fig7} (b). 
The median surface brightness of QSFs 
is $4.21\times  10^{-16}$ ergs cm$^{-2}$ s$^{-1}$ arcsec$^{-2}$.
Table~\ref{tbl-3} summarizes the \feii\ parameters of QSFs.

QSFs are dense shocked clumps and, as we discussed in \S~3.1,   
their 1.64 $\mu$m emission should be almost entirely due to the \feii\ line. 
In radiative atomic shocks, \feii\ line brightness is roughly proportional to 
$n_0 v_s^3$ where $n_0$ is pre-shock density and $v_s$ is  
shock speed \citep[e.g., see Figure~\ref{fig:fig6} of ][]{koo16}. 
If the shocks propagating into the QSFs are driven by 
hot gas at constant ram pressure $(n_0v_s^2)$, 
then the \feii\ brightness of QSFs is expected to increase 
with $v_s$ up to $v_s\sim 300(n_0/10^3)^{1/3.4}$~\kms\ at which the cooling 
time becomes comparable to the age of Cas A 
\citep[e.g., see eq. 36.33 of][]{draine11}.
\cite{mck75} visually examined the 
brightnesses of QSFs in the 
H$\alpha$ image of \cite{van71} and claimed that indeed the 
QSFs with largest shock speeds are fainter.  
According to our result, however, 
there is no correlation between the \feii\ brightness and 
radial velocity of QSFs.
In Figure~\ref{fig:fig8}, we plot extinction-corrected 
\feii\ surface brightness versus radial velocity ($\vrad$) of QSFs for those with radial velocities known from optical observations \citep{ala14}.
For example, the QSFs with the highest radial velocities 
($\vrad \simlt -350$ km s$^{-1}$ or 
$\vrad \simgt +100 $ km s$^{-1}$) are not fainter 
nor brighter than the other QSFs.
Instead, the radial velocities of the 
brightest QSFs are $\sim -100$~\kms.
It has been known that QSFs are systematically blueshifted \citep{van85,law95,ala14}.  
The \feii-line flux-weighted mean radial velocity is $-120$~\kms. 
It is possible that this velocity represents a     
systemic motion of QSF knots, but the subtraction of such systemic 
velocity does not improve the correlation.   

A difficulty in examining  
Figure~\ref{fig:fig8} is the projection effect. 
We may simply assume that the shocks propagating into QSFs 
are mainly in the radial direction from the SNR geometrical center, 
in which case the observed radial velocities 
would be the line-of-sight component of the total speed of a shock. 
\cite{mck75} derived the total shock speeds of individual QSFs assuming that 
QSFs are located near the forward shock front. 
This was justifiable because at that time
the thickness of the region ($\Delta R_s$) between the ambient and 
reverse shocks 
was thought to be 10-15\% of the SNR radius; since QSFs are likely to be   
quickly destroyed once swept up by the dense shocked ejecta, 
they may be assumed to be in the region between the ambient and 
reverse shocks. 
We now have a better estimate for $\Delta R_s$, and it is a considerable 
fraction of the distance to the SNR radius ($R_s$), 
i.e., $\Delta R_s/R_s \approx (2.5~{\rm pc}-1.7~{\rm pc})/2.5~{\rm pc}=0.32$ \citep{ree95,got01}. 
So the QSFs that we see can spend  
$\sim 100$~yr, or the corresponding fraction of the SNR age, 
embedded in the region of hot, shocked gas
(e.g., see Figure 10 of \citealt{nozawa10}).
For comparison, the disruption timescale for 
a dense clump of radius $a$ by a shock wave of speed $v_s$ is 
$\sim 2a/v_s=200(a/0.01~{\rm pc})(v_s/100~{\rm km~s}^{-1})$~yr \citep{mck75}.
So QSFs with radius larger than $\sim 0.005$~pc are likely to survive until they 
encounter the ejecta shell where they will be destroyed. 
Since 0.005 pc is $1/3$ of the median radius of QSFs, 
although the initial radii of QSFs could be significantly less than their observed radii, 
we may consider that the QSFs that we see will mostly survive until they encounter the ejecta shell
and that their lifetime is $\sim 100$~yr or even longer considering the time that they spend 
in the ejecta shell.  This is consistent with our conclusion in \S~5.1 that the lifetime of QSFs
is $\simgt 60$~yr. 
Therefore, 
in principle the QSFs outside the reverse shock can be anywhere along the line of sight and 
the correction factor for the projection can be very large. 
For example, the bright QSFs with small radial velocities, e.g.,   
Knots 33, 30, and 40 in Figure~\ref{fig:fig8}, might be 
moving almost perpendicularly on the plane of sky, although they could be bright because 
they are interacting with shocked SN ejecta (see below).  
For the QSFs inside the reverse shock, one can derive 
the lower and upper limits to shock speeds, but, considering the 
complex structure of shocked clumps, these limits will be also quite uncertain.

 In Figure~\ref{fig:fig8}, the red solid lines show the expected \feii\ brightness 
for shocks of constant ram pressure of $10^6$ and $10^7$~cm$^{-3}$ (\kms)$^{2}$. 
The ambient shock in Cas A is probably expanding into 
CSM with $n(r)\propto r^{-2}$ density distribution,  
and the pre-shock density at the current location of the shock front is about 1 cm$^{-3}$ 
\citep{leejj14b}. 
The shock speed is about 5000~\kms, so
the thermal pressure just behind the ambient shock is 
$n_0 v_s^2\approx 2.5 \times 10^7$~cm$^{-3}$ (\kms)$^2$. 
The pressure drops behind the shock front \citep[e.g.,][]{chevalier82}, so
the pressure driving a shock into QSFs in the shocked ambient medium 
will be  somewhat lower, e.g., $\sim 2\times 10^7$ cm$^{-3}$ (\kms)$^2$.
Figure~\ref{fig:fig8} shows that most QSFs are fainter than predicted by shock models  
by an order of magnitude, except the ones on the main ejecta shell.
We attribute this to the small area-filling factor of 
\feii\ emitting region in QSFs. 
The morphology of clumps  
disrupted by blast wave seen in numerical simulations is quite complex 
\citep[e.g.][]{klein94}.
The HST images also show 
filamentary and clumpy structure of QSFs \citep{ala14}. 
The \feii\ emission is likely to be emitted 
from such dense filaments/fragments filling a small 
fraction of the QSFs' geometrical areas in Table~\ref{tbl-1}. 
The lack of correlation between the \feii\ brightness and 
$\vrad$ could be also due to the complex spatial and kinematic 
structure of QSFs. The shocks are probably propagating into QSFs in all directions and 
the observed radial velocities might be intensity-weighted 
averages of their line-of-sight velocity components. 
In some QSFs, 
the radial velocity varies more than 100~\kms\ over the structure.
Therefore, the radial velocities cannot be a good tracer of shock speeds.
One thing to notice in Figure~\ref{fig:fig8} is that the QSFs on the main ejecta shell
are relatively bright, e.g., Knots 33, 30, 40, 48, and 1.  
This is not just because we used a higher threshold for the detection (see \S~4.3) 
since there are no bright ($\ge 10^{-14}$ erg cm$^{-2}$ s$^{-1}$ sr$^{-1}$) 
QSFs detected outside the main ejecta shell. 
We suggest that this is because those QSFs are indeed physically  
interacting with the main ejecta shell; 
when QSFs are swept-up by dense ejecta shell, a strong shock will 
be driven into QSFs by the shell's large ram pressure, so that they will be brighter.    
A detailed spectral mapping of individual QSFs 
can be helpful to understand how the \feii\ brightness is related to their 
physical properties.  

We can derive the total mass of Fe atoms ($\mfe$) from   
the \feii\ 1.644 $\mu$m luminosity ($\lumfeii$) by
\begin{equation}
\mfe=2.93\times 10^{-6} \left(\frac{\lumfeii}{L_\odot} \right)
\left(\frac{\ffep}{1.0} \right)^{-1}
 \left(\frac{\flevel}{0.01} \right)^{-1}\quad M_\odot
\end{equation} 
where $\ffep$ is the fraction of Fe in Fe$^+$ and 
$\flevel$ is the fraction of Fe$^+$ in the level a$^4$D$_{7/2}$.
We used Einstein coefficient $A_{21}(=5.07\times 10^{-3}$~s$^{-1}$)
of \cite{deb11} and the collision strength of \cite{ramsbottom07}. 
In statistical equilibrium, 
$\flevel=0.0054$--0.01 at temperatures 5,000--10,000 K and  
electron density $1\times 10^4$ cm$^{-3}$.
A median electron density of QSFs is 
a few times $10^4$ cm$^{-3}$ \citep{leeyh17}, and  
the characteristic temperature of the \feii\ 1.644 \mum\ line
emitting region is $T=7000$~K \citep{koo16}. So we may take $\flevel\approx 0.01$.  
The total \feii\ 1.644 $\mu$m flux of QSF knots is 
$7.47 \times 10^{-12}$ erg cm$^{-2}$ s$^{-1}$, so that 
$\lumfeii=2.70$~\lsun (Table~\ref{tbl-2}).
This implies $\mfe\approx 7.9\times 10^{-6}$~\msun, and 
for the cosmic abundance of Fe, i.e., $X({\rm Fe})=3.5\times 10^{-5}$, 
hydrogen+helium mass $M({\rm H+He})\approx 0.23 M_\odot$.
For comparison, \cite{van71} obtained 
0.025~\msun\ from their H$\alpha$ flux applying an  
extinction correction factor of 100. 
If we consider that the extinction to Cas A is $A_V=6$--11 mag and 
that the QSFs in the western area which contribute 
$\simgt 60$\% of the total \feii\ line flux (see below) are not seen in H$\alpha$, 
the two estimates roughly agree with each other.  
For comparison, the total mass of the X-ray emitting 
swept-up material, representing the smooth red supergiant (RSG) wind,   
is $\sim 6~M_\odot$ \citep{leejj14b}. 
Hence, according to our result, the mass of {\em visible} QSFs 
is about $4$\% of the total swept-up wind mass. 
If we naively scale the observed mass using 
the ratio of the total volume of the SNR to 
the volume of the region between ambient and reverse shocks (see \S~5.2), 
the fraction becomes a little larger ($\sim 6$~\%).

\subsection{QSFs and Mass Loss History of the Cas A Progenitor}

Figure~\ref{fig:fig6} 
shows some distinct features of the spatial distribution of QSFs. 
The most prominent one is 
the arc structure in the south (hereafter the `southern QSF arc'),
which had been noticed by previous studies in H$\alpha$+\nii\ images.  
\cite{law95} noticed that QSFs along this arc structure, 
together with the QSF knots in the northern area,  
are located along the edges of $4.'7\times 2.'3$ ellipse with 
P.A. of $25\arcdeg$.
They noted that this QSF distribution, however, 
cannot be an inclined circle or an ellipse centered at the expansion center 
because all QSFs have velocities between 
$-100$ and $-200$ \kms\ and because there is 
no correlation between radial velocities and the P.A..
The southern QSF arc is also prominent in the [\ion{N}{2}] $\lambda$6583 image 
of \cite{ala14}. The measured velocities cover a wider range, e.g., +110~\kms\ to $-180$ \kms, 
but show no systematic variation (see Figure~\ref{fig:fig6}).

The QSFs along the southern QSF arc seem to have appeared 
over some considerable period of time.  
\citet{van83} and \cite{van85} were the first to identify QSFs 
in this area; Knots 10, 17, 24, and 25 
(= R36, R40, R37, and R38 in \citealt{van85}; see 
Figure \ref{fig:fig9new}).\footnote{\citet{van83} and \cite{van85} detected another faint knot  
in the southwestern corner of Figure \ref{fig:fig9new}. 
They classified this knot (R39) as QSF because 
it was not seen in either \stwo- or \oiii-sensitive 
plates. But later, this knot was found to have 
a large proper motion, i.e., $0\farcs65\pm0\farcs10$,  
and reclassified as a N-rich fast-moving flocculus (FMF) 
by \citet{fes87}. A faint knot matching R39 
is visible in our \deepimage, although it is not marked in Figure \ref{fig:fig9new}.} 
For the bright Knot 24 (= R37),   
they noted that it can be seen in all available plates 
from 1951 to 1980. So this QSF had appeared before 1951 and 
has been visible during $\simgt 62$ yrs including our \deepimage.
For Knot 10 (= R36), they noted that it had been invisible until 1958 
and marginally visible in 1965 and clearly seen on all plates taken since 1967.
In a recent \halpha+\nii\ image shown in Figure~\ref{fig:fig9new}, 
we can confirm these QSFs.  The \halpha+\nii\ image 
in Figure~\ref{fig:fig9new} has been produced from the the Isaac Newton Telescope (INT) 
Wide Field Survey (WFS)  data obtained in 2005 October.\footnote{The data are available at 
the INT WFS archive site http://apm3.ast.cam.ac.uk/cgi-bin/wfs/dqc.cgi.}
On the other hand, we note that Knot 7 in our image 
is not visible in any of the previously published H$\alpha$+\nii\ images 
including the 2009 \nii\ image of \citet{ala14} (see also Figure~\ref{fig:fig9new}),
so that this bright knot (and possibly the surrounding faint knots too) appeared after 2009. 
If we assume that the moment that they appear is the moment that  
these dense knots are swept-up by the SNR shock, then 
Knot 7 might be slightly further away, e.g., by $\sim 0.3$ pc 
using the SNR shock speed of 5000~\kms,  
from the SNR center than Knot 24 although the former appears closer on the sky.
And the knots along the southern QSF arc, together with the 
QSF knots in the northern area that \citet{law95} noticed, 
are likely distributed along an elliptical ring, 
but appear as they are because of projection. 
The fact that these knots are systematically blue-shifted  
suggests that the elliptical ring is likely located on the near side 
of the SNR shock.

Figure \ref{fig:fig6} reveals another prominent feature, namely, the cluster of QSFs in the 
western area beyond the main ejecta shell.
Figure~\ref{fig:fig10} shows the spatial and flux distribution 
of QSFs as a function of 
P.A. measured from north to east with respect to the explosion center.
It shows that QSFs are concentrated in the southwestern area 
near P.A.$\sim 260\arcdeg$ or toward the direction of the SW counterjet. 
The QSFs between P.A.=$240\arcdeg$ and $280\arcdeg$  
contribute $\simgt 60$\% of the extinction-corrected flux of QSFs.
The median value of their proper motions is $0.02$ arcsec yr$^{-1}$, 
which gives an expansion timescale of $\sim 6,000$~yr but with a large 
uncertainty ($\pm 6,000$~yr) due to our short time interval. 
For comparison, the expansion timescale of QSFs from optical studies is   
$11,000\pm2,000$~yr \citep{kam76,van85}.
The expansion timescale, however, could be longer if the observed proper motion is partly 
due to shock motion into QSFs (see below).
Therefore, if the QSFs that we see were ejected during a short time interval, 
Figure \ref{fig:fig10} suggests that there was an eruptive mass 
loss $\simgt 10,000$ years ago, mainly to the west. 

It has been suggested that QSFs are dense clumps embedded in a smooth 
RSG wind from the Cas A progenitor \citep{che03}. 
High-resolution studies of nearby RSGs indeed reveal inhomogeneities 
indicating anisotropic and episodic mass ejection   
\citep{mauron97,dewit08,ohnaka17}. 
But we note that the proper motion  
($0\farcs02$ yr$^{-1}$) of QSFs corresponds to 320~\kms, which is much greater than the 
typical expansion speed of RSG wind, i.e., $\sim 15$ \kms\ \citep{mauron11}. 
Therefore, if QSFs are dense clumps in RSG wind, the observed 
proper motion (and radial velocities) should be almost entirely due to the shock motion, and 
the expansion time scale for the QSFs  
needs to be $\simlt (2.5~{\rm pc})/(15~{\rm km~s}^{-1})\sim 1.6\times 10^5$~yr. 
Based on stellar evolution models, \citet{yoon10} suggest that, at about this time 
before the SN explosion, 
RSG stars with initial mass of $\simgt 17$~\msun\ have a brief period of `superwind' phase 
where the star experiences strong H envelope pulsation 
with an order-of-magnitude enhanced mass loss rate.   
The mass ejection could 
have been asymmetric if  a non-radial mode of pulsation  
was dominant. 
On the other hand, if the progenitor was in a binary system, 
a dense and clumpy spiral structure can be formed   
in the equatorial plane by wind-wind interaction \citep[e.g., see Figure~\ref{fig:fig1} of ][]{pan15}, 
which will become a circular or an elliptical 
arc at a distance of $\simgt 1$ pc from the explosion center 
at the time of SN explosion. If the arc is inclined, it may appear as 
the southern QSF arc. \cite{law95} ruled out such possibility because 
all QSFs have negative velocities and 
there is no correlation between the radial velocity and P.A.. 
But if the progenitor system was moving through the interstellar space 
at $\sim -100$~\kms\ and the velocity structures of individual QSFs 
are dominated by shocks, then it does not seem to be impossible to 
see the observed velocity distribution of QSFs.  

Another possible explanation for QSFs is that  
they are dense clumps ejected in the common envelope
phase of the progenitor binary system.
The small ejecta mass together with N-rich CSM   
suggests that the progenitor was a He star in a binary system \citep{young06}. 
The non-detection of the ex-companion implies that the 
companion star could have been either white dwarf or neutron star 
\citep{kerzendorf17}.   
A plausible evolutionary scenario for such system 
is that the progenitor in the late RSG phase expands  
and forms a common envelope through the Case C mass transfer to the companion.
Then there could be a brief and 
explosive mass loss in the common envelope phase. 
For example, the outburst of `luminous red nova (LRN)' is 
considered to be the  mass loss from a binary system in 
its common envelope phase \citep{ivanova13}.
The characteristic expansion velocities of the LRN outburst 
are 200--1000~\kms\ \citep[][and references therein]{ivanova13}, 
so that the dense clumps ejected during the outburst will 
propagate ballistically through the slow ($\sim 15$~\kms) 
RSG wind because of large density contrast. 
The `large' proper motion (320~\kms) of QSFs 
can be explained by such outburst and their systematic blueshift by 
one-sided ejection toward us, although there seems to be 
no plausible explanation for such highly asymmetric ejection. 

\section{Fast-Moving Knots and Shocked Dense SN Ejecta}\label{sec:fmk}

\subsection{FMKs outside the Main Ejecta Shell}\label{subsec:fmkhst}


We have identified 179 FMKs outside the main ejecta shell 
(Table~\ref{tbl-1}).
Their measured proper motions range from 0.18 to 0.76 arcsec yr$^{-1}$ with a median of 0.49 arcsec yr$^{-1}$ or 7900~\kms. 
They are mostly found in four regions:  
(1) NE jet, (2) southeastern area of 
${\rm P.A.}=80^\circ$--$150^\circ$, (3) SW counterjet, 
and (4) northern area of  ${\rm P.A.}=320^\circ$--$5^\circ$.
(see Figure~\ref{fig:fig1}).     
No FMKs are found in broad regions between these areas. 
That spatial distribution is similar to the distribution of optical ejecta 
knots reported by \citet{fes06} and \citet{ham08}.
From the HST observations in various filters, 
1825 outlying optical ejecta knots have been found and they are 
divided into N-rich, O-rich, and S-rich knots \citep{ham08}.
These chemically distinct ejecta knots show very different spatial distributions; 
N-rich knots are arranged in a broad shell with gaps in the north and south, 
O-rich knots which are fewer in number are found in limited P.A. and 
radial distance ranges, and S-rich knots are mainly concentrated in the NE jet, SW counterjet, and in the southeastern area (\citealt{fes06}; \citealt{ham08}; see also \citealt{mil13} and \citealt{fes16}). 
The FMKs in the \deepimage\ are mostly   
S-rich knots: among the 128 outlying FMKs 
located in the area covered by \citet{ham08},\footnote{The other 51 FMKs are located near the main ejecta shell which was not explored   
by \cite{ham08}. For example, the northern area where we have found 46 FMKs 
was not included in their study. 
Note that, in this paper, the 
term `FMKs' refers to any fast-moving ejecta knots, while in \citet{ham08}  
the term is used either for ejecta knots in the main ejecta shell or 
for outlying ejecta knots chemically similar to the `main-shell FMKs'.}
6 (5\%), 2 (2\%), and 93 (73\%) knots are 
N-rich, O-rich, and S-rich ejecta knots in the catalog of \cite{ham08}, 
respectively.   
Twenty seven (21\%) FMKs do not have counterparts in the catalog of 
\cite{ham08}. 
In Figure~\ref{fig:fig11}, we show the \deepimage\  
of the NE jet area, southeastern area, and SW jet area together with the HST F098M image.  
The HST F098M image was obtained in 2011 November 18 \citep{fes16}, 
1.9 years earlier than the \deepimage,
and the FMK contours are shifted considering their proper motions.  
We can see that essentially all FMKs have counterparts in the HST image 
including the ones without counterparts in the catalog of \citet{ham08}.  
(The HST F098M image has a higher sensitivity than 
the HST F850LP used by \citealt{ham08}.) 
N-rich knots of \cite{ham08} are marked by plus symbols, and 
only a few N-rich knots have counterparts in our \deepimage.
In particular, in the southeastern area, there are numerous N-rich knots 
but none of them have counterparts in \feii\ emission. 
Some knots could have been missed because 
of our narrow bandwidth ($\pm 2600$~\kms), 
but according to spectroscopic observations \citep{fesen01a}, 
N-rich knots with radial velocities outside our band 
are expected to be rare in this area. 
Therefore, we tentatively conclude that most FMKs in the \deepimage, including the ones 
not cataloged by \cite{ham08}, are S-rich optical ejecta knots.

In Figure \ref{fig:fig11}, we notice that the outer FMKs in the southeastern area, 
e.g.,  Knots 240, 241, 262, 271, and 273,  
are generally brighter in the \deepimage\ than in the HST F098M image.
Figure~\ref{fig:fig12} shows this quantitatively, where 
we plot the flux ratio ($\fluxhubble/\fluxfeii$) of FMKs 
as a function of angular distance from the explosion center in three areas. 
The left frame is for the observed fluxes while the 
right frame is for the extinction-corrected fluxes. 
The extinction correction
raises the flux ratios systematically, but the trend remains the same. 
Note  that 
$\fluxfeii$ is the flux within the narrow band ($\pm 2600$~\kms), so there could be 
missing flux compared to the broadband F098M flux.  
Outlying ejecta knots often show radial velocities as large as 5000~\kms, although   
the majority of knots have more modest radial velocities, e.g.,  
$-3000$ to +1000~\kms\ \citep{fes96, fesen01a, mil13}.
With that caveat in mind, we note 
in Figure~\ref{fig:fig12} that FMKs at angular distance $\simgt 2\farcm5$ in the southeastern  
area (green squares) have relatively low  $\fluxhubble/\fluxfeii$. 
The missing \feii\ flux will shift the data points  
upwards not downwards. So these knots appear to be Fe rich, indicating that 
a significant fraction of Fe ejecta may have mixed out to large radius in the southeast.
It is interesting that this is the area where the 
diffuse X-ray Fe-K emission extends beyond the main ejecta shell 
\citep{hwa04}. \citet{del10} proposed an ejecta piston model where 
Fe ejecta moving faster than the average ejecta is pushing the O-burning layer above.  
Dense Fe-rich knots embedded in diffuse Fe-rich ejecta appear to be consistent with 
small scale clumping and large scale anisotropy seen
in the recent three-dimensional numerical simulations of neutrino-driven 
SN explosions \citep{won15,orl16}.  
The line ratio derived from the two images of several years' interval,
however, can be affected by the brightness changes or flickering during the interval \citep{fes11}, so we need NIR spectroscopy to confirm such 
chemical variation of the  
outlying FMKs. That will be done in our forthcoming paper.

\subsection{\feii\ Emission from FMKs}

FMKs are shocked ejecta material and most of the 1.64 \mum\ emission from FMKs is 
probably due to \feii\ 1.644 \mum\ line (see \S~\ref{subsec:prelude}). 
FMKs are  much fainter than QSFs, e.g., by an order of magnitude in flux (Table~\ref{tbl-3}). 
Their brightness, i.e., flux divided by the area, is also less than that of QSFs  by a factor of 
2--3. That is due to a tendency of the shocked ejecta
to cool so quickly that the much of the Fe remains in higher ionization
states in the region that produces \feii\ emission in normal abundance shocks.  
The total observed and extinction-corrected fluxes of FMKs are  
$1.1\times 10^{-13}$~erg cm$^{-2}$ s$^{-1}$
and $4.6\times 10^{-13}$~erg cm$^{-2}$ s$^{-1}$, respectively.

There is considerable uncertainty about the fractions of
various FMK emission lines that are produced in the shocked
gas and the pre-shock photoionization region 
\citep{itoh81, sutherland95, blair00}.
\cite{docenko10}, analyzing {\it Spitzer} and ISO maps, 
found that the emission from singly ionized species such 
as \stwo\ and \feii\ arises mostly from the post-shock
photoionized region.
The densities of around $3\times 10^4$ cm$^{-3}$ obtained from the \feii\ lines 
by \cite{koo13} also imply that the \feii\ line mostly comes from 
the post-shock region.  We note that \cite{docenko10} 
obtained a density an order of magnitude higher for the FMKs
based on the [\ion{O}{1}] FIR line ratio from ISO.  One possible reason
for this discrepancy may be that the [\ion{O}{1}] lines can form in
much colder gas than the \feii\ lines because of their low
excitation potential, and the density may be higher there.
Another possibility is that \feii\ is suppressed at the high
densities where the [\ion{O}{1}] lines are produced, so emission
from intermediate densities dominates.  The third, and in
our view most likely, possibility is that the lines come from
both pre-and post-shock gas, and their line ratios are in the
low- and high-density limits, respectively.  In that case, the
observed ratio will apparently indicate an intermediate density,
while in fact it indicates the fractions of emission from the
pre-shock and post-shock gas.  The fairly high densities given
by the \feii\ line ratio indicate that most of the \feii\ 
emission is from the post-shock region, in agreement with the
conclusions of \cite{docenko10}.  This also fits in
with the \stwo\ observations of \cite{ala14}, who found
that the \stwo \ ratio varies all the way from the low density
limit to the high density limit, with most positions giving a
value near the high density limit. 

\subsection{\feii\ Luminosity of Shocked Dense Ejecta}\label{subsec:felum}

The total \feii\ 1.644 \mum\ luminosity of Cas A in the \deepimage\ is 
$2.3$~\lsun. If we apply extinction-correction to each pixel 
of the main shell in the same way as we did for the knots, we  
obtain 11.8~\lsun\ for the extinction-corrected \feii\ 1.644 \mum\ luminosity of Cas A.
The QSFs contribute 23\% of this (Table~\ref{tbl-3}), so that the total luminosity due to    
the shocked SN ejecta in Cas A is 9.1~\lsun, most of which is from the main ejecta shell.
Note that this does not include the  
fast-moving ($>2600$~\kms) ejecta undetected in our narrow-band image, 
but their contribution should be almost negligible. 
For comparison,  the observed \feii\ 1.644 \mum\ luminosity of the Kepler SNR is  
0.7 \lsun\ \citep{oli89}, so Cas A is much brighter than Kepler. In the Galaxy, however,  
there are many SNRs interacting with dense environment and very bright in \feii\ emission
(Lee, Y.-H. et al. in preparation), so that Cas A is not a particularly bright SNR in \feii\ emission. 
What makes Cas A special is that, in Cas A, the \feii\ emission is mostly from the shocked SN ejecta not from the shocked circumstellar or shocked interstellar medium.

We can estimate the total mass of Fe SN ejecta associated with the \feii\ emission.
If we substitute the extinction corrected luminosity (9.1~\lsun) of shocked SN 
ejecta in equation (1), we obtain an Fe mass of $ 2.7\times 10^{-5}$~\msun.  
Essentially all of this mass is in the main ejecta shell.
This is in fact a lower limit because there could be neutral Fe atoms that 
have already cooled down.  However, because the optical emission has been
steadily brightening in recent decades \citep{patnaude14}, it is likely that
the amount of material in shocked ejecta that have faded is small compared to
the present \feii-line emitting mass.  There may also be some Fe in higher 
ionization states, and a significant fraction of Fe may remain in dust.
\citet{micelotta16} find that only about 10\% of the dust in FMKs survives,
but much of the destruction occurs after the grains move from the ejecta
knots into the hotter surrounding material.
For comparison, the mass of diffuse, X-ray emitting Fe 
ejecta has been estimated to be  $\sim 0.1$~\msun \citep{hwa12}, while
the mass of unshocked Fe ejecta in the interior 
inferred from the $^{44}$Ti emission is also $\sim 0.1$~\msun \citep{gre14}.
Hence, the mass of dense \feii-line emitting 
Fe ejecta appears only a few $10^{-4}$ of the total Fe ejecta mass.

\section{Conclusion and Summary}\label{sec:conclusion}

Cas A is one of the first SNRs detected in the optical band  \citep{baa54,min59}, and since then it 
has been a subject of extensive optical studies. Studies in the NIR band, however, have been  limited.
Here we present a deep \feii+\sii\ image of Cas A 
obtained 
with a narrow-band filter centered at 1.644 $\mu$m emission. 
The image gives an unprecedented panoramic view of Cas A, 
showing the prominent main shell and FMKs of shocked SN ejecta, compact QSFs of shocked CSM, 
and faint diffuse emission from the unshocked SN ejecta in the interior, all in one single image.
The emission from shocked CSM and probably that from the SN ejecta too are  
mostly \feii\ 1.644 $\mu$m emission, while 
the emission from the unshocked SN ejecta is likely dominated by \sii\ 1.645 $\mu$m emission.
The image reveals for the first time the 
detailed structure of the interior diffuse emission closely related to the explosion dynamics 
and that of the western portion of the main ejecta shell disrupted by the jet. 
The image also shows a comprehensive sample of QSFs that can be used to infer  
the mass-loss history of the progenitor system.
The results presented in this paper  
show that the NIR band can provide 
new and important information about the final evolution and SN explosion of massive stars. 
In the following, we summarize our main results.

\bigskip
\noindent
(1) The \deepimage\ shows in detail the complex structure of the diffuse 
emission in the interior of Cas A from the unshocked SN ejecta. 
Some emission features have unique morphology implying their close association with SN 
explosion dynamics,  e.g., the Eastern Arc surrounding the explosion center and   
the pillars protruding from the main ejecta shell and pointing to the explosion center. 
The overall morphology, however, is not correlated with $^{44}$Ti emission but 
well correlated with \siii\ 34.81 $\mu$m emission, so that the emission is likely due to  
the \sii\ 1.645 $\mu$m line, not the \feii\ 1.644 $\mu$m line. 

\noindent
(2) The \deepimage\ reveals the disrupted western part of the main ejecta shell that was not 
visible in the optical band due to large extinction.   
The area is filled with bright and faint clumps of $\simlt 20''$.
The faint clumps, in general, show large proper motions over 5 years,
and they are probably the fragments of the disrupted ejecta shell. 
In contrast, the bright clumps in general show little proper motion, and 
they are probably QSFs of circumstellar origin. 
Some large QSFs appear to be interacting with shocked diffuse ejecta material, 
implying that they have been swept-up by the reverse shock very recently.
The possibility that the anomalous radio properties in this area could  
be due to the dense CSM and/or the shell fragments needs to be explored.

\noindent
(3) We have identified 309 knot features in the \deepimage\ and derived their 
geometrical parameters and \feii\ fluxes.  
We classified their nature, i.e., QSF vs FMK, by comparing with another 1.64 \mum\ image of Cas A 
obtained 5 years earlier in 2008 August and also with the HST images sensitive to 
\stwo\ lines. 
One hundred thirty knots are classified as QSFs and the remaining 179 knots are classified as FMKs.
A catalog of 309 knots is presented in Table~\ref{tbl-1} and     
their characteristic \feii\ parameters are summarized in Table~\ref{tbl-3}.

\noindent
(4) We have searched for counterparts of QSFs in 
H$\alpha$/\nii$\lambda\lambda$6548, 6583 images published since 1951 or vice versa.
Eighty-eight (68\%) of 130 QSFs in the \deepimage\ are newly identified.  
On the other hand, most QSFs seen in previous studies  
are visible in our \deepimage, indicating that the lifetime 
of QSFs is $\simgt 60$~yr. Most QSFs are likely to survive until they encounter the ejecta shell.
For QSFs with radial velocities known from optical observations, 
we explored whether the radial velocity can be an indicator of shock speed.  
We found that the \feii\ 1.644~\mum\ brightness is not correlated with the radial velocity and that it is  
much lower than what shock models predict. We attribute this to the complex 
spatial and velocity structure of QSFs. 

\noindent
(5) The total H+He mass of visible QSFs  is estimated $\approx 0.23 M_\odot$, which 
corresponds to 4\% of the total mass of the swept-up CSM. The fraction of dense clumps in the RSG wind  
could be a little larger (6\%) if we consider the finite lifetime of QSFs. 
The spatial distribution of QSFs is highly asymmetric. 
In addition to the prominent southern arc known from optical observations,  
the \deepimage\ revealed a cluster of 
QSFs in the western area, contributing  more than 60\% of the total flux. 
The QSF distribution indicates that the progenitor system ejected its envelope eruptively  
to the west just before the explosion, i.e., $10^4$ to $10^5$ yr before the explosion depending on
whether these QSFs are ejected at $\sim 15$~\kms\ in RSG phase 
or ejected at a few 100~\kms\ in common envelope phase of the progenitor binary system. 

\noindent
(6) We have identified 179 FMKs outside the main ejecta shell in the \deepimage.
FMKs are generally much fainter than QSFs in \feii\ emission.  
Most of these FMKs are S-rich optical ejecta knots, and we can identify the counterparts of essentially all FMKs in the HST F098M image. We note that the FMKs outside the southeastern disrupted ejecta shell are 
relatively brighter in the \deepimage\ than in the HST F098M image
compared to the other FMKs.
This is the area where the the diffuse X-ray Fe K emission extends  
beyond  the main ejecta shell, and the FMKs there may be Fe-rich. 
That needs to be confirmed from spectroscopic observations. 

\noindent
(7) The total extinction-corrected \feii\ 1.644 \mum\ luminosity of Cas A is 11.8~\lsun, 
23\% of which is from QSFs. 
The total Fe luminosity of shocked SN ejecta is 9.1~\lsun\ and it is 
mostly from the main ejecta shell. 
The inferred \feii-emitting, dense Fe mass in the ejecta shell is 
$2.7\times 10^{-5}$~\msun, which is only a few times $10^{-4}$ of the total Fe ejecta mass. 

\acknowledgments
We thank the referee for highly perceptive comments, 
which helped us improve the discussion about 
QSFs and FMKs significantly.  
We also thank Rob Fesen for many helpful comments.
This research was supported by Basic Science 
Research Program through the National Research Foundation of
Korea(NRF) funded by the Ministry of Science, ICT and future Planning (2017R1A2A2A05001337).



\vspace{5mm}
\facilities{UKIRT, Hale, \spitzer, HST, INT}

\software{IDL}



{}



\newpage

\newpage

\newpage
 
\newpage


\newpage
\clearpage

\newpage
\begin{figure}[ht]
\begin{center}
\includegraphics[scale=0.2]{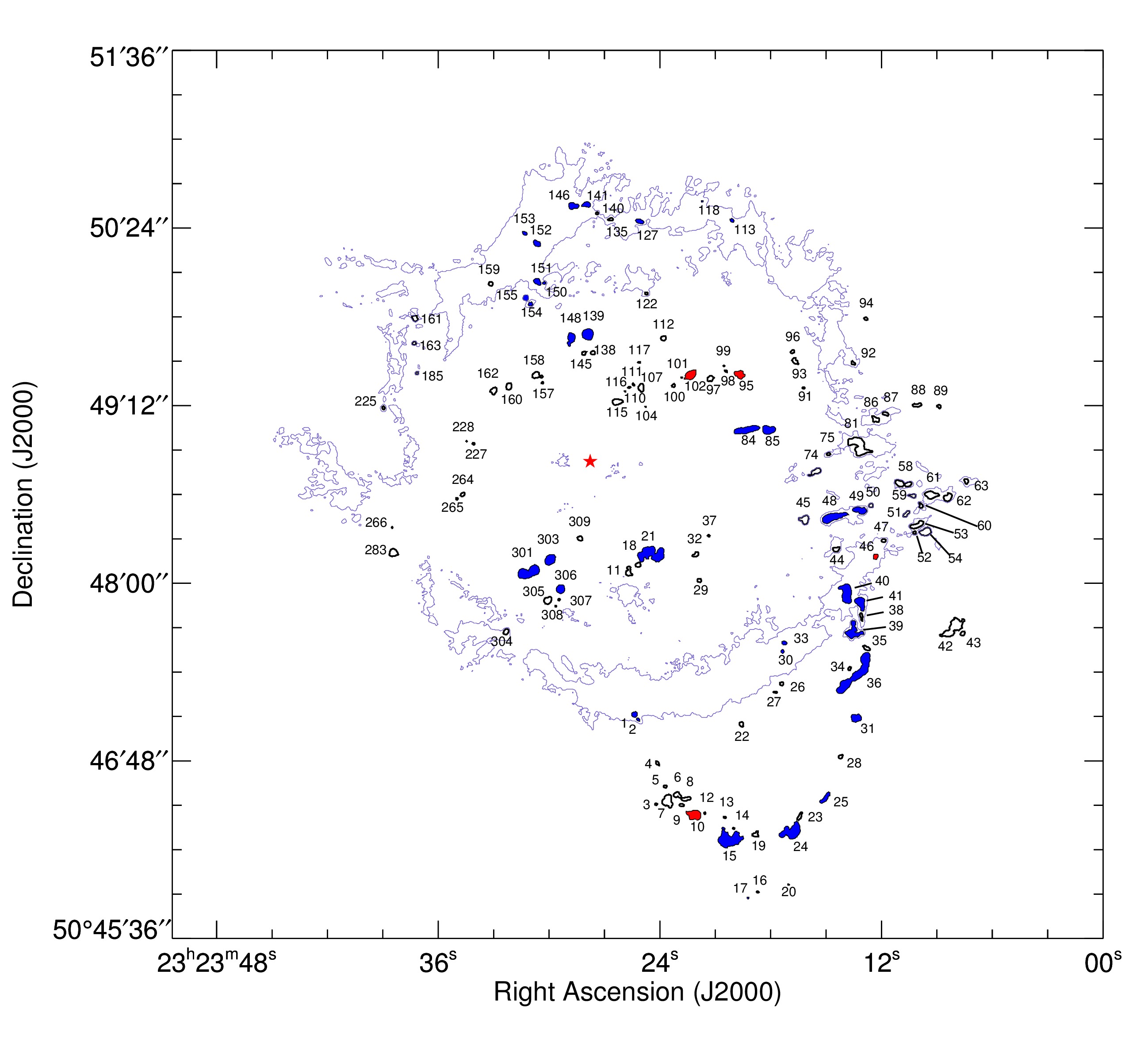}
\caption{
Finding chart of QSFs in Table~\ref{tbl-1}. The filled symbols with blue and red colors  
represent QSFs previously detected in \nii\ $\lambda$6583 line and 
with negative and positive radial velocities, respectively \citep{ala14}.  
The background blue contour is to show the main ejecta shell in the \deepimage\ and it is at  
$4 \times 10^{-18}$~\bunit. The red star symbol represents the 
explosion center. 
\label{fig:fig6}
}
\end{center}
\end{figure}
\clearpage

\newpage
\begin{figure}[ht]
\begin{center}
\includegraphics[scale=0.15]{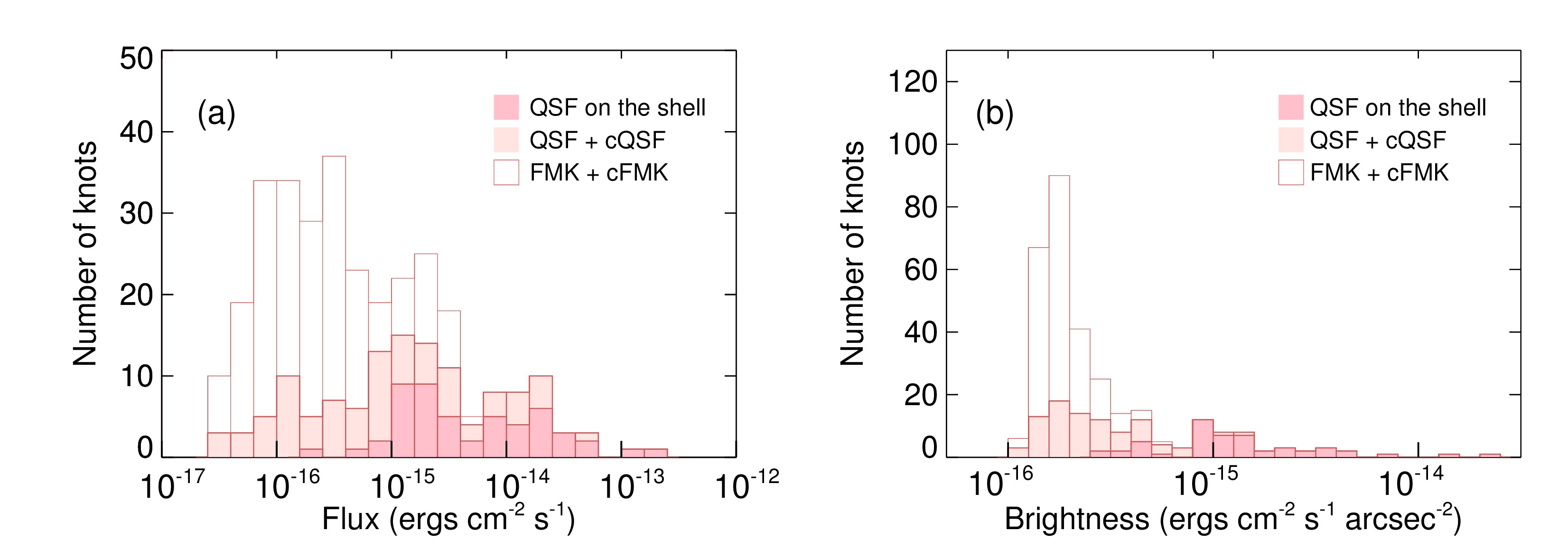}
\caption{Distribution of observed flux (a) and brightness (b) of knots identified in the \deepimage. 
\label{fig:fig7}
}
\end{center}
\end{figure}
\clearpage

\newpage
\begin{figure}[ht]
\begin{center}
\includegraphics[scale=0.22]{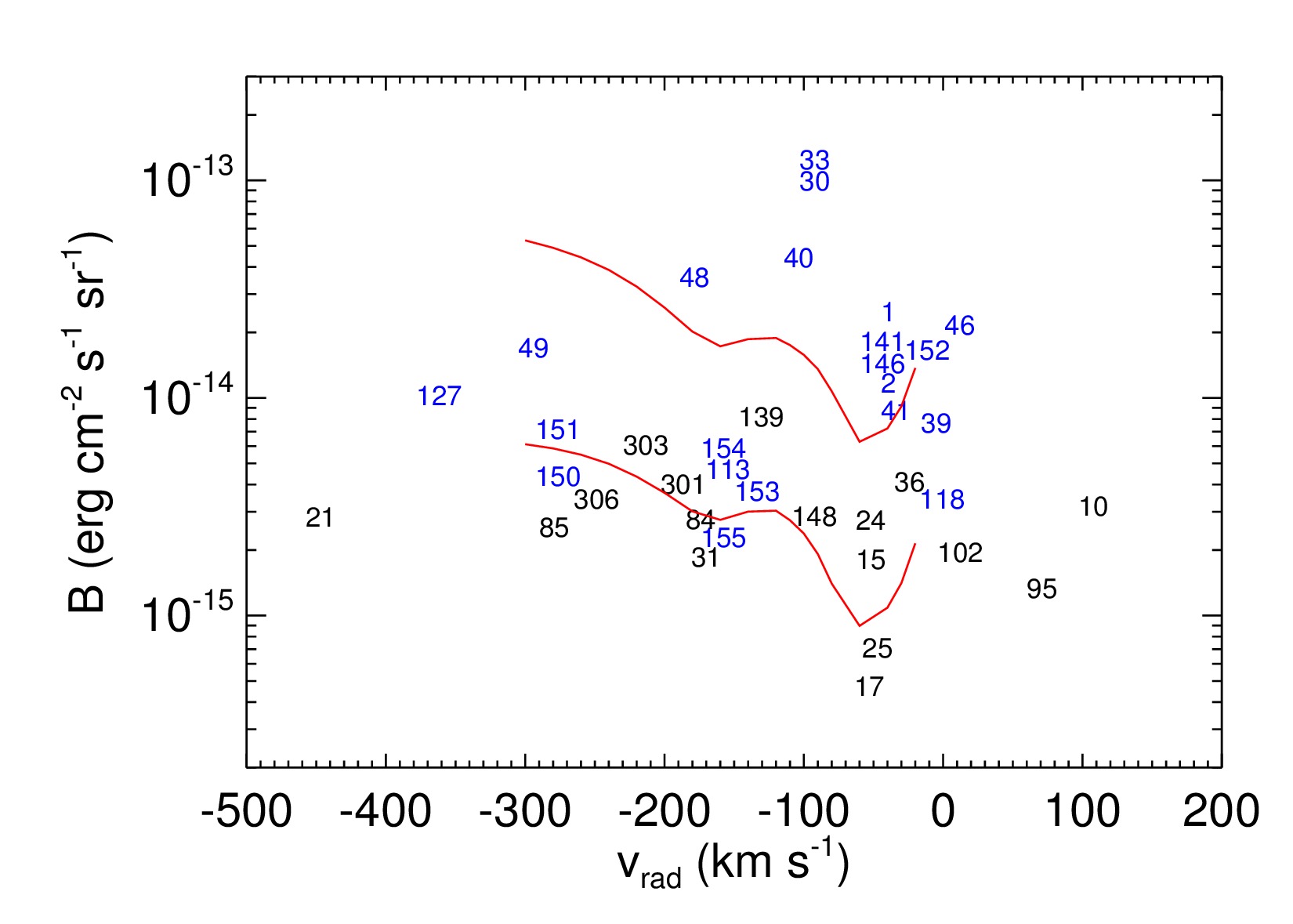}
\caption{\feii\ 1.644 \mum\ brightness vs.   
observed radial velocity for QSFs with known radial velocities. 
The radial velocities are from \cite{ala14}. 
QSFs are labeled by their IDs in this paper (see Figure~\ref{fig:fig6}).
The QSFs of blue label are the ones on the main shell. 
The red solid lines show the brightness from shock models 
for shock speeds equal to $|v_{\rm rad}|$. 
The lower and upper curves are for constant ram pressure of 
$n_{\rm ion}v_s^2=10^6$ and $10^7$ ~cm$^{-3}$ (\kms)$^2$, respectively. 
See Appendix~\ref{sec:appendix} for the other shock parameters. 
\label{fig:fig8}
}
\end{center}
\end{figure}
\clearpage

\newpage
\begin{figure}[ht]
\begin{center}
\includegraphics[scale=0.35, trim=300 0 0 0]{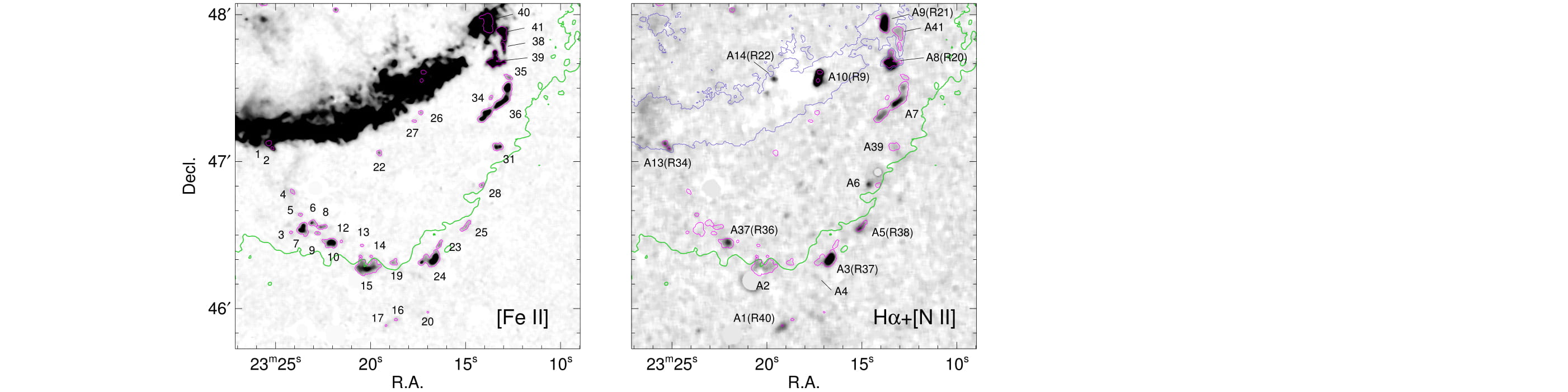}
\caption{(Left) Enlarged view of the southern QSF arc. 
QSFs are marked by magenta contours and labeled by their IDs in Table 1 (see also Figure~\ref{fig:fig6}).
The green contour marks the outer boundary of the SNR in radio.
(Right) The same area in \halpha+\nii\  emission. 
The image has been produced  from the INT WFS data obtained in 2005 October and a 
continuum is subtracted by using an $r^{\prime}$ broad-band image.  
QSFs identified in optical emission are labeled 
by their IDs in the catalogs of \cite{ala14} and \cite{van85} (see Table 2).
\label{fig:fig9new}
}
\end{center}
\end{figure}
\clearpage

\newpage
\begin{figure}[ht]
\begin{center}
\includegraphics[scale=0.3]{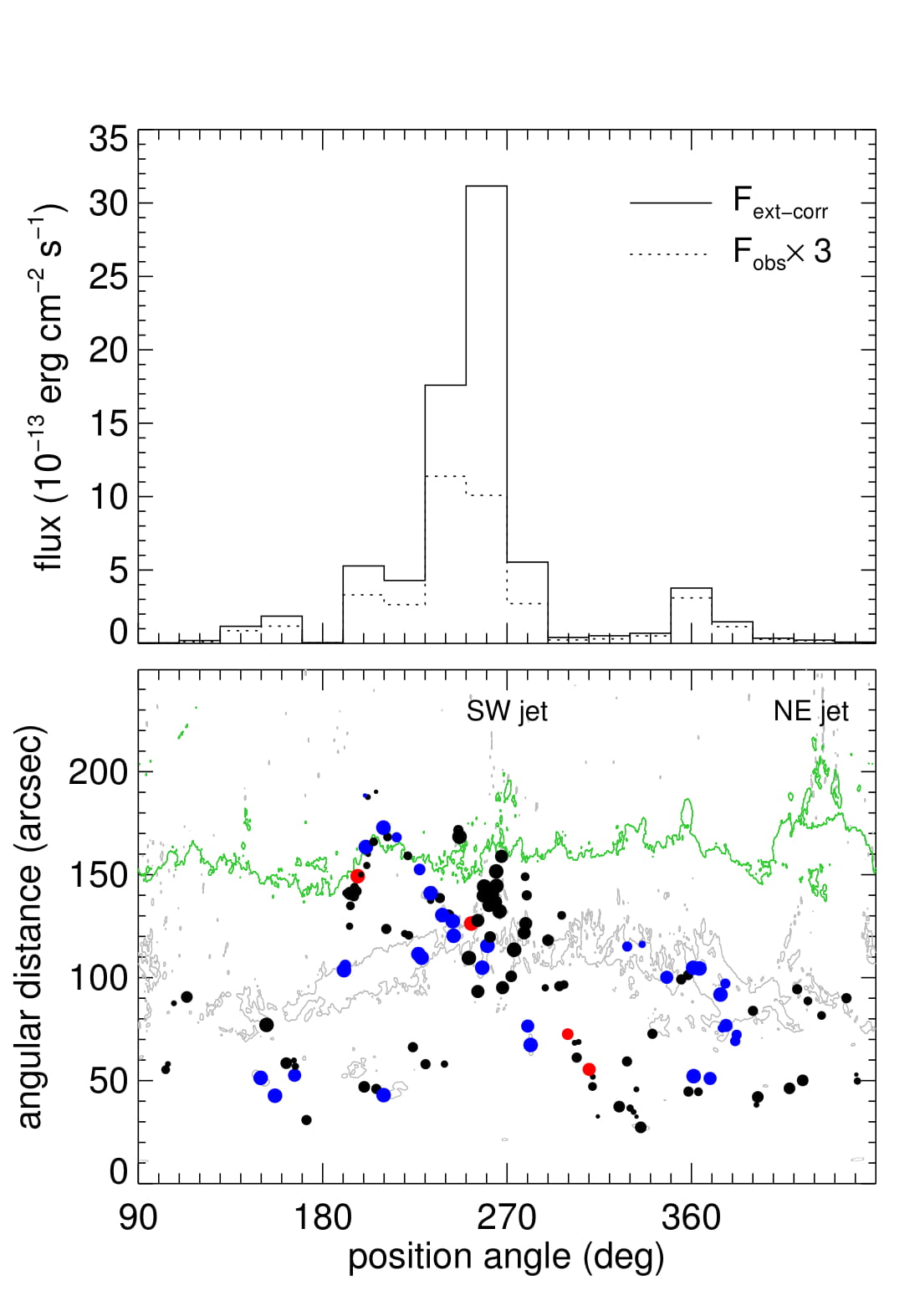}
\caption{Top: Fluxes of QSFs versus their position angle (P.A.).
The P.A. is measured from north to east with respect to the explosion center, so
the west, for example, corresponds to P.A.$=270\arcdeg$. 
The dotted histogram shows the distribution of observed fluxes multiplied by three while
the solid histogram shows the distribution of extinction-corrected fluxes.
Bottom: Angular distance of QSFs measured from the explosion center versus their P.A..
Solid circles represent the locations of QSFs. 
The area of the symbols is proportional to the QSF flux while the color of the symbols   
represents the QSF velocities, i.e., ${\rm blue}={\rm negative~velocities}$, ${\rm red}={\rm positive~velocities}$, and ${\rm black}={\rm unknown~velocities}$.
The grey contour shows the location of the main ejecta shell.   
\label{fig:fig10}
}
\end{center}
\end{figure}
\clearpage

\newpage
\begin{figure}[ht]
\begin{center}
\includegraphics[scale=0.3]{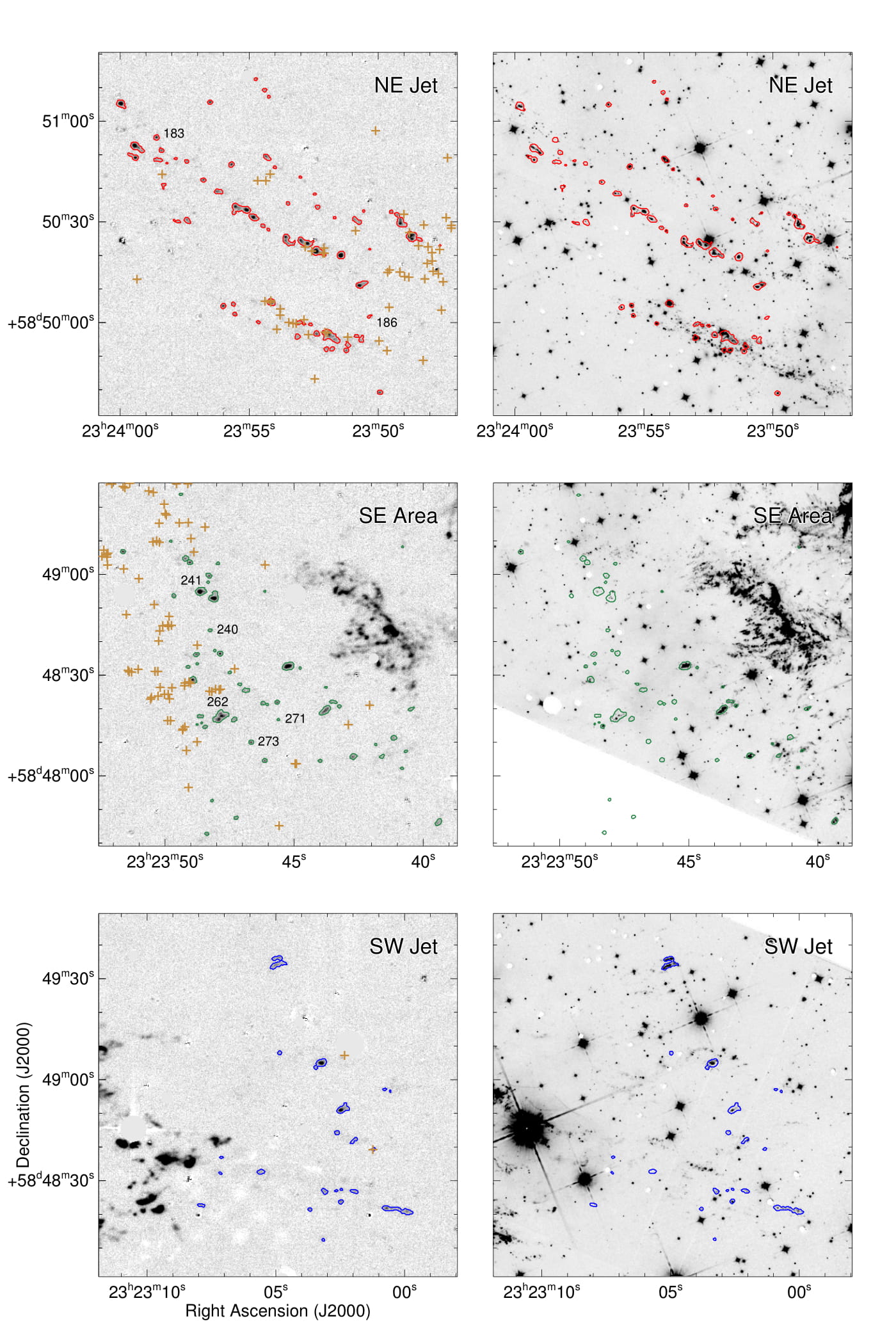}
\caption{
Enlarged view of NE jet, SE area, and SW counterjet 
in the \deepimage\ (left) and in the HST F098M image (right).
The contours in the \deepimage\ mark the identified FMKs, while    
those in the HST F098M image are  the same contours but shifted to the 
position at the epoch (2011 November) of the HST image by assuming that 
the FMKs have been freely expanding from the explosion center since A.D. 1670. 
FMKs with 
$F_{\rm F098M}/F_{\rm [Fe II] 1.644 \mu m}\le 0.3$ 
in Figure~\ref{fig:fig12} are labeled by their IDs.
The brown plus symbols mark the positions of N-rich optical 
ejecta knots in the catalog of \cite{ham08}. 
\label{fig:fig11}
}
\end{center}
\end{figure}
\clearpage

\newpage
\begin{figure}[ht]
\begin{center}
\includegraphics[scale=0.18]{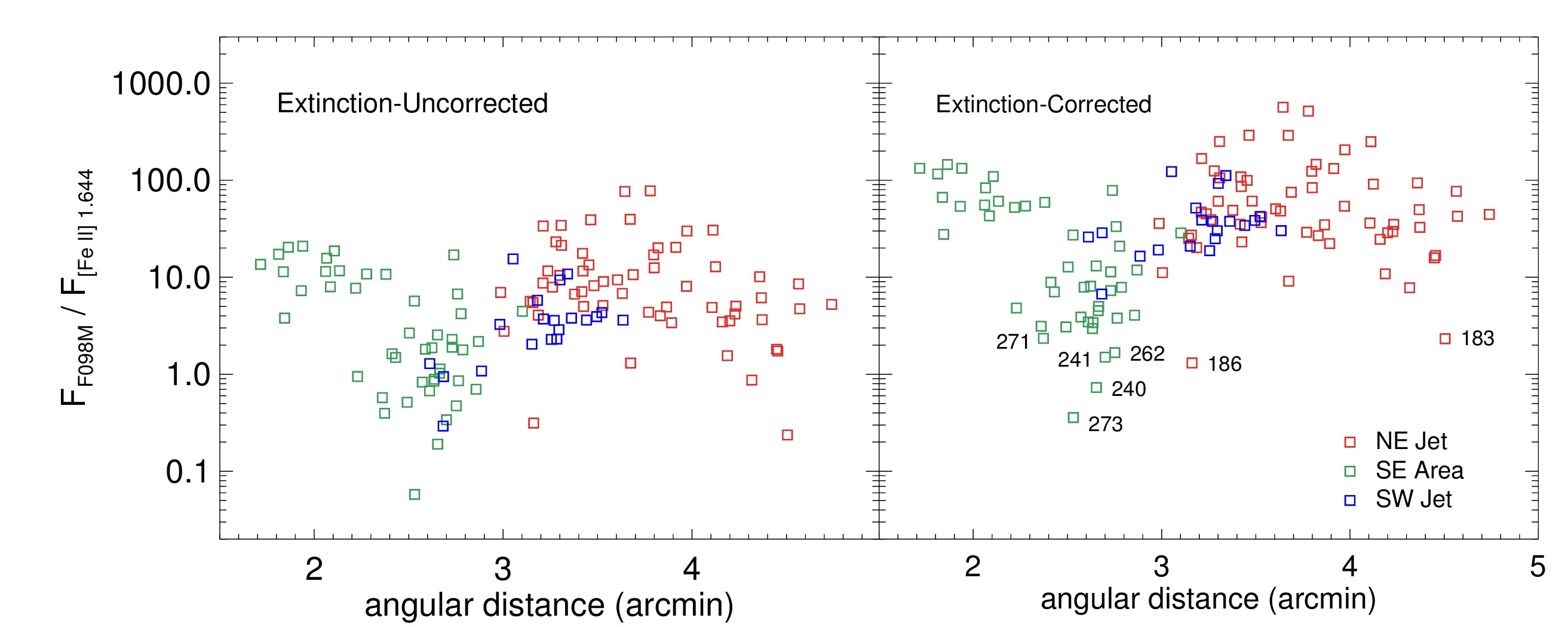}
\caption{
Left: Flux ratio $F_{\rm F098M}/F_{\rm [Fe II] 1.644 \mu m}$ of FMKs  
with the 
observed \feii\ 1.644 \mum\ flux $\ge 5\times 10^{-17}$ erg cm$^{-2}$ s$^{-1}$ (see Figure~\ref{fig:fig8}) 
as a function of angular distance from the explosion center.
Right: Same as left but with the interstellar extinction corrected. 
FMKs with 
$F_{\rm F098M}/F_{\rm [Fe II] 1.644 \mu m}\le 0.3$ 
are labeled by their IDs.
The $1\sigma$ uncertainties in the observed and extinction-corrected flux ratios are 
typically $\sim 4$~\% and $\sim 15$~\%, respectively.  
\label{fig:fig12}
}
\end{center}
\end{figure}
\clearpage

\newpage
\startlongtable
\begin{deluxetable*}{rllrrrrccccr}
\tabletypesize{\scriptsize}
\setlength{\tabcolsep}{3pt}
\tablecaption{Catalog of knots in the deep [\ion{Fe}{2}$]+[$\ion{Si}{1}] image.\label{tbl-1}}
\tablewidth{0pt}
\tablecolumns{13} \tablehead{
\colhead{ID} & \colhead{$\alpha$(J2000)} & \colhead{$\delta$(J2000)} &
\colhead{$D_{\rm max}$} & \colhead{$D_{\rm min}$} & \colhead{$\psi_{\rm ellipse}$} &
\colhead{Area} & \colhead{$F_{\rm obs}$} & \colhead{$F_{\rm ext-corr}$} &
\colhead{$\mu$} & \colhead{$\theta_\mu$} &\colhead{Class} \\
\colhead{ } & \colhead{ } & \colhead{ } & \colhead{(arcsec)} &
\colhead{(arcsec)} & \colhead{(deg)} & \colhead{(arcsec$^2$)} &
\colhead{(erg cm$^{-2}$ s$^{-1}$)} & \colhead{(erg cm$^{-2}$ s$^{-1}$)} &
\colhead{($\arcsec$ yr$^{-1}$)} & \colhead{(deg)} & \colhead{ } 
}
\decimalcolnumbers
\startdata
1\tablenotemark{\it s}         & 23:23:25.362 &  58:47:07.47 &   2.33 &   2.07 &   107 &   3.77 &    1.62E-14 &    8.91E-14 &     0.03 &    171.2 & QSF\tablenotemark{\it o}         \\
2\tablenotemark{\it s}         & 23:23:25.165 &  58:47:05.42 &   1.46 &   0.88 &    48 &   1.01 &    2.14E-15 &    1.23E-14 &     0.02 &    102.1 & QSF\tablenotemark{\it o}         \\
3\phm{\tablenotemark{\it s}}   & 23:23:24.196 &  58:46:31.07 &   1.20 &   0.99 &   100 &   0.93 &    1.69E-16 &    1.04E-15 &  \nodata &  \nodata & cQSF\phm{\tablenotemark{\it o}}  \\
4\phm{\tablenotemark{\it s}}   & 23:23:24.115 &  58:46:47.71 &   2.30 &   1.22 &    39 &   2.15 &    3.23E-16 &    1.74E-15 &  \nodata &  \nodata & cQSF\phm{\tablenotemark{\it o}}  \\
5\phm{\tablenotemark{\it s}}   & 23:23:23.699 &  58:46:38.27 &   1.61 &   1.23 &    89 &   1.54 &    3.92E-16 &    1.99E-15 &     0.01 &     90.0 & QSF\phm{\tablenotemark{\it o}}   \\
6\phm{\tablenotemark{\it s}}   & 23:23:23.040 &  58:46:34.80 &   3.53 &   2.34 &    73 &   5.95 &    1.86E-15 &    9.01E-15 &     0.09 &   -100.2 & QSF\phm{\tablenotemark{\it o}}   \\
7\phm{\tablenotemark{\it s}}   & 23:23:23.565 &  58:46:32.33 &   5.40 &   4.37 &    11 &  17.10 &    9.13E-15 &    4.42E-14 &     0.04 &    -21.2 & QSF\phm{\tablenotemark{\it o}}   \\
8\phm{\tablenotemark{\it s}}   & 23:23:22.567 &  58:46:33.26 &   4.44 &   1.61 &    98 &   5.51 &    1.68E-15 &    9.24E-15 &     0.02 &   -113.2 & QSF\phm{\tablenotemark{\it o}}   \\
9\phm{\tablenotemark{\it s}}   & 23:23:22.804 &  58:46:30.73 &   2.23 &   1.24 &    84 &   2.19 &    4.58E-16 &    2.22E-15 &  \nodata &  \nodata & cQSF\phm{\tablenotemark{\it o}}  \\
10\phm{\tablenotemark{\it s}}  & 23:23:22.124 &  58:46:26.73 &   5.98 &   3.79 &    82 &  16.97 &    9.35E-15 &    5.14E-14 &     0.06 &     99.9 & QSF\tablenotemark{\it o}         \\
\enddata
\tablecomments{(1) Knot ID; (2)--(3) central coordinate from ellipse fitting; 
(4)--(5) major and minor axis from ellipse fitting; 
(6) P.A. of the major axis measured from north to east; 
(7) area measured from contours; (8) observed flux; (9) extinction-corrected flux; 
(10) proper motion of knots; (11) P.A. of proper motion measured from north to east; (12) classification of knots, with a prefix `c' referring to a candidate.
}
\tablenotetext{\it s}{Knots (QSFs) superposed on the main ejecta shell (see \S~\ref{subsec:knotshell}).}
\tablenotetext{\it o}{QSFs with optical counterparts from \citet{ala14} and \citet{van85} (see Table~\ref{tbl-2}).}
\tablecomments{{\it Knots 16, 17, 20}: No counterpart in 2008, but by comparing with \citet{van83, van85} and \cite{ala14}, they are classified as QSF candidates. {\it Knot 18}: No counterpart in 2008 but there is a counterpart without a positional shift in the HST F098M; this can be a newly-appearing QSF. {\it Knots 55, 64, 65, 66}: Near the bright QSFs in the western, disrupted ejecta shell, but classified as FMK candidates because they are well aligned with the SW jet direction from the explosion center. The HST F098M image shows no emission features probably because of large extinction. {\it Knots 91, 93, 96, 227, 228}: The HST F098M image shows emission features but without a positional shift between the HST and our deep [\ion{Fe}{2}$]+[$\ion{Si}{1}] image; thus they are classified as QSF candidates. {\it Knots 287, 298, 299, 300}: Outside of the HST F098M image but the counterparts with proper motions are confirmed in the HST 2004 image, so they are classified as FMKs.}
\tablecomments{Table 1 is published in its entirety in the machine-readable format.
      A portion is shown here for guidance regarding its form and content.}
\end{deluxetable*}
\clearpage

\newpage
\startlongtable
\begin{deluxetable*}{rllcccccc}
\tabletypesize{\scriptsize}
\setlength{\tabcolsep}{5pt}
\tablecaption{QSFs with optical counterparts\label{tbl-2}}
\tablewidth{0pt}
\tablecolumns{6} \tablehead{
\colhead{ } & \colhead{ } & \colhead{ } &
\multicolumn{3}{c}{\citet{ala14}} & \multicolumn{3}{c}{\citet{van85}} \\
\cline{4-6} \cline{7-9}
\colhead{ID} & \colhead{$\alpha$(J2000)} & \colhead{$\delta$(J2000)} &
\colhead{ID$_{\rm opt}$} & \colhead{$v_r$} & \colhead{$v_{r,err}$} &
\colhead{ID$_{\rm opt}$} & \colhead{$\mu$} & \colhead{$v_r$} \\
\colhead{ } & \colhead{ } & \colhead{ } &
\colhead{ } & \colhead{(km s$^{-1}$)} & \colhead{(km s$^{-1}$)} &
\colhead{ } & \colhead{($\arcsec$ yr$^{-1}$)} & \colhead{(km s$^{-1}$)} 
}
\decimalcolnumbers
\startdata
1\tablenotemark{\it s}         & 23:23:25.362 &  58:47:07.47 &      A13 & -35                            &       10 &        R34 &     0.01 &  \nodata \\
2\tablenotemark{\it s}         & 23:23:25.165 &  58:47:05.42 &      A13 & -35                            &       10 &        R34 &     0.01 &  \nodata \\
10\phm{\tablenotemark{\it s}}  & 23:23:22.124 &  58:46:26.73 &      A37 & 107                            &       13 &   R36n,s &     0.04 &  \nodata \\
15\phm{\tablenotemark{\it s}}  & 23:23:20.210 &  58:46:16.89 &       A2 & -53                            &        9 &   \nodata &  \nodata &  \nodata \\
16\phm{\tablenotemark{\it s}}  & 23:23:18.688 &  58:45:55.48 & \nodata & \nodata                        &  \nodata &        R40 &     0.02 &  \nodata \\
17\phm{\tablenotemark{\it s}}  & 23:23:19.213 &  58:45:53.03 &       A1 & -54                            &        9 &        R40 &     0.02 &  \nodata \\
21\phm{\tablenotemark{\it s}}  & 23:23:25.027 &  58:48:12.10 &      A31 & -448\tablenotemark{\dag}       &       13 &  R1c,e,w &     0.02 &      432 \\
24\phm{\tablenotemark{\it s}}  & 23:23:16.766 &  58:46:19.15 &       A3 & -53                            &        2 &   R37n,s &     0.04 &  \nodata \\
25\phm{\tablenotemark{\it s}}  & 23:23:15.011 &  58:46:33.56 &       A5 & -48                            &        8 &        R38 &     0.08 &  \nodata \\
30\tablenotemark{\it s}        & 23:23:17.328 &  58:47:33.00 &      A10 & -93                            &        2 &     R9c,s &     0.02 &     -128 \\
31\phm{\tablenotemark{\it s}}  & 23:23:13.333 &  58:47:05.97 &      A39 & -171                           &       37 &   \nodata &  \nodata &  \nodata \\
33\tablenotemark{\it s}        & 23:23:17.230 &  58:47:36.41 &      A10 & -93                            &        2 &        R9n &     0.01 &     -198 \\
36\phm{\tablenotemark{\it s}}  & 23:23:13.338 &  58:47:23.89 &       A7 & -25\tablenotemark{\dag}        &        8 &   \nodata &  \nodata &  \nodata \\
39\tablenotemark{\it s}        & 23:23:13.473 &  58:47:41.08 &       A8 & -6\tablenotemark{\dag}         &       10 &        R20 &     0.02 &  \nodata \\
40\tablenotemark{\it s}        & 23:23:13.856 &  58:47:56.51 &       A9 & -105\tablenotemark{\dag}       &       29 &        R21 &     0.02 &  \nodata \\
41\tablenotemark{\it s}        & 23:23:13.089 &  58:47:52.57 &      A41 & -35                            &       28 &   \nodata &  \nodata &  \nodata \\
46\tablenotemark{\it s}        & 23:23:12.264 &  58:48:11.41 &      A11 & 11                             &       12 &        R35 &     0.17 &  \nodata \\
48\tablenotemark{\it s}        & 23:23:14.574 &  58:48:27.39 &      A12 & -180\tablenotemark{\dag}       &        7 &        R19 &     0.01 &     -182 \\
49\tablenotemark{\it s}        & 23:23:13.111 &  58:48:30.15 &      A38 & -295                           &       21 &        R19 &     0.01 &     -182 \\
84\phm{\tablenotemark{\it s}}  & 23:23:19.274 &  58:49:02.83 &      A29 & -175                           &       17 &        R23 &     0.03 &  \nodata \\
85\phm{\tablenotemark{\it s}}  & 23:23:18.074 &  58:49:02.80 &      A44 & -280                           &       18 &        R23 &     0.03 &  \nodata \\
95\phm{\tablenotemark{\it s}}  & 23:23:19.645 &  58:49:25.25 &      A15 & 70                             &       13 &   \nodata &  \nodata &  \nodata \\
102\phm{\tablenotemark{\it s}} & 23:23:22.308 &  58:49:25.10 &      A36 & 6                              &       22 &   \nodata &  \nodata &  \nodata \\
112\phm{\tablenotemark{\it s}} & 23:23:23.784 &  58:49:39.98 & \nodata & \nodata                        &  \nodata &      (R14) &     0.17 &     -340 \\
113\tablenotemark{\it s}       & 23:23:20.057 &  58:50:27.70 &      A28 & -161                           &       16 &   \nodata &  \nodata &  \nodata \\
118\tablenotemark{\it s}       & 23:23:21.677 &  58:50:35.46 &      A27 & -7                             &        7 &     R11,R12 &     0.02 &     -134 \\
127\tablenotemark{\it s}       & 23:23:25.075 &  58:50:27.36 &      A33 & -368                           &       24 &        R17 &     0.02 &     -283 \\
139\phm{\tablenotemark{\it s}} & 23:23:27.895 &  58:49:41.56 &      A18 & -137                           &        2 &     R2n,s &     0.02 &     -208 \\
141\tablenotemark{\it s}       & 23:23:27.963 &  58:50:34.17 &      A25 & -50                            &        4 &        R10 &     0.02 &      -20 \\
146\tablenotemark{\it s}       & 23:23:28.687 &  58:50:33.62 &      A25 & -50                            &        4 &       R29e &     0.01 &      -45 \\
148\phm{\tablenotemark{\it s}} & 23:23:28.808 &  58:49:39.83 &      A17 & -99                            &        3 &         R3 &     0.01 &     -163 \\
150\tablenotemark{\it s}       & 23:23:30.271 &  58:50:02.34 &      A23 & -283                           &       10 &   \nodata &  \nodata &  \nodata \\
151\tablenotemark{\it s}       & 23:23:30.635 &  58:50:02.89 &      A23 & -283                           &       10 &   \nodata &  \nodata &  \nodata \\
152\tablenotemark{\it s}       & 23:23:30.664 &  58:50:18.40 &      A22 & -18                            &        2 &     R7n,s &     0.01 &      -66 \\
153\tablenotemark{\it s}       & 23:23:31.325 &  58:50:22.52 &      A24 & -140                           &       18 &        R26 &     0.04 &  \nodata \\
154\tablenotemark{\it s}       & 23:23:31.012 &  58:49:53.83 &      A19 & -164                           &        3 &     R5e,w &     0.02 &     -265 \\
155\tablenotemark{\it s}       & 23:23:31.265 &  58:49:56.38 &      A19 & -164                           &        3 &   \nodata &  \nodata &  \nodata \\
161\tablenotemark{\it s}       & 23:23:37.291 &  58:49:48.03 & A40 & \nodata                        &  \nodata &        R28 &     0.02 &  \nodata \\
225\tablenotemark{\it s}       & 23:23:39.000 &  58:49:11.71 & \nodata & \nodata                        &  \nodata &        R13 &     0.00 &     -170 \\
301\phm{\tablenotemark{\it s}} & 23:23:31.102 &  58:48:05.07 &      A16 & -193                           &        6 &     R4c,e &     0.02 &     -162 \\
303\phm{\tablenotemark{\it s}} & 23:23:29.942 &  58:48:10.20 &      A35 & -220                           &       58 &        R4w &     0.01 &     -170 \\
306\phm{\tablenotemark{\it s}} & 23:23:29.374 &  58:47:58.31 &      A34 & -255                           &       31 &        R25 &     0.02 &  \nodata \\
\enddata
\tablecomments{(1) Knot ID; (2)--(3) central coordinate from ellipse fitting; 
(4)--(6) optical QSF ID and radial velocity with error from \citet{ala14}; 
(7)--(9) optical QSF ID, proper motion, and radial velocity from \citet{van85}---if several optical QSFs are matched with a  [\ion{Fe}{2}] knot, we use their mean proper motion and velocity.}
\tablenotetext{\it s}{Knots (QSFs) superposed on the main ejecta shell (see Section~4.1.3).}
\tablenotetext{\dag}{For QSFs with subknot structures in \citet{ala14}, we use the mean velocity of the subknots.}
\tablecomments{ {\it Knot 112}: \citet{van85} obtained a large proper motion and 
put parenthesis on the ID, i.e., ``(R14)'', to indicate a need for confirmation. 
But the large proper motion is likely due to confusion with another fragment in the area. 
The proper motion we estimate by comparing with the 2008 image is 0.02~$\arcsec$ yr$^{-1}$.
}
\end{deluxetable*}
\clearpage

\begin{deluxetable}{l c c}
\tablecaption{ [\ion{Fe}{2}] 1.644 \mum\ Properties of QSFs and FMKs\label{tbl-3}}
\tablewidth{0pt}
\setlength{\tabcolsep}{8pt}
\tablecolumns{3} \tablehead{
\colhead{Parameter} & \colhead{QSF} & \colhead{\hfil FMK \hfil} 
}
\startdata
 Number of identified knots\tablenotemark{a}      &          130  	              &            179 \\
 Geometrical radius (arcsec)	& 1.03 (0.28--4.39)   &    0.61 (0.28--2.18) \\ 
 $D_{\min}/D_{\max}$				& 0.70 (0.21--0.98)   &  0.72 (0.14--1.00)\\ \\
\underline{Observed} & & \\
Flux ($10^{-15}$ ergs cm$^{-2}$ s$^{-1}$)  & 1.86 (0.044--239)    & 0.24 (0.039--5.23) \\
Brightness ($10^{-16}$ ergs cm$^{-2}$ s$^{-1}$ arcsec$^{-2}$)   &    	4.21 (1.36--245)  	& 	2.08 (1.35--6.38)  	\\
Total flux ($10^{-12}$ ergs cm$^{-2}$ s$^{-1}$)  & 1.27 & 	0.11 \\ \\
\underline{Extinction corrected} & & \\
Flux ($10^{-15}$ ergs cm$^{-2}$ s$^{-1}$)  &  9.12 (0.14--1125)  & 1.08 (0.12--22.1)\\
Brightness ($10^{-16}$ ergs cm$^{-2}$ s$^{-1}$ arcsec$^{-2}$)   &   20.5 (4.51--1188) 	  	& 	 	8.94 (4.85--26.0) \\
Total flux ($10^{-12}$ ergs cm$^{-2}$ s$^{-1}$)  &  7.47 & 	0.46  \\
Luminosity (\lsun) & 2.70 & 0.17 
\enddata 

\tablenotetext{a}{These numbers include 36 candidates for QSF and 12 candidates for FMKs. 
See \S~\ref{subsec:knotclass} for an explanation of candidates.}
\tablecomments{The quoted values are the median and the numbers in parenthesis are the minimum and maximum.}
\end{deluxetable}
\clearpage

\newpage
\appendix

\section{\sii\ 1.645 $\mu$\MakeLowercase{m} Line from Radiative Shocks}\label{sec:appendix}

\sii\ 1.645 $\mu$m line is not included in most shock models, and there is little 
recent atomic data. Here we compute the brightness of the \sii\ line   
in radiative atomic shock with QSF and FMK abundances.
For the collision strength of \sii, we use the result of \cite{pindzola77}.  
Because \ion{Si}{1} is neutral, the excitation cross section goes to zero at 
threshold (0.781 eV=9,060 K for the $^1D_2$ level), which means that the 
(velocity-averaged) collision strength is very small at temperatures below about 4000 K. 
For comparison, the excitation energy of the \feii\ 1.644 \mum\ line is 11400 K, but because
\ion{Fe}{2} is an ion, the collision strength does not go to zero at threshold, and
the excitation rate is higher at a few thousand K.
Excitation of the \sii\ 1.645 $\mu$m line requires both the presence of \ion{Si}{1} 
($T < 10000$ K) and a high temperature enough to overcome the Boltzmann
factor ($T > 2000$ K). We scale \ci\ 9850 \AA\ line flux to estimate the \sii\ brightness.
The \ci\ 9850 \AA\ line is similar to the \sii\ 1.645 $\mu$m line 
in the behavior of its excitation rate with temperature and 
in the ionization fraction of \ion{C}{1}. 
The \sii\ 1.645 $\mu$m emission could also come from  recombination of \ion{Si}{2} ion.
The existence of \ion{Si}{2} at low temperatures in the post-shock layer 
guarantees that there will be some \sii\ emission produced by recombination.  To compute the contribution of recombination to the 
\sii\ line we use the recombination coefficients of \citet{abdel-naby12}
and assume that 1/4 of the recombinations go by way of the singlets and that all those pass through the $^1D$ state of the ground level.

For QSF, we adopt the solar abundances of \cite{asplund2009} except He and N 
which are assumed to be enhanced by a factor of 2. 
Note that the solar abundances of Si and Fe are comparable, i.e., 
$X({\rm Si/H})=3.55\times 10^{-5}$ and $X({\rm Fe/H})=3.47\times 10^{-5}$ by number, 
respectively. The shock code that we use is the one
developed by \citet{raymond79} with updated atomic parameters \citep{koo16}.
We have run models with shock speed  
$v_s = 50$--300~\kms\ and pre-shock density 
$n_0 = 10^2$--$3\times 10^3$ cm$^{-3}$, which are the characteristic parameters of QSFs 
\citep{mck75,che03}. Magnetic field strength is fixed to 5 $\mu$G. 
We find that the \sii\ line is only about  $\le 2$\% 
as strong as the \feii\ line in these models. 
This is mainly because \ion{Fe}{2} is the majority ion of iron in the
photoionized, radiative zone in shocked gas, while \ion{Si}{1} is a minority species there
\citep[see also Figure~\ref{fig:fig6} of][]{koo16}.   
The emission from the recombination of \siii\ ions is not significant either.  
Therefore, for QSFs, the 1.64 $\mu$m 
line should be dominated by the \feii\ line emission.

%

%
%

\begin{figure}[b!]
\centering
\includegraphics[width=100mm]{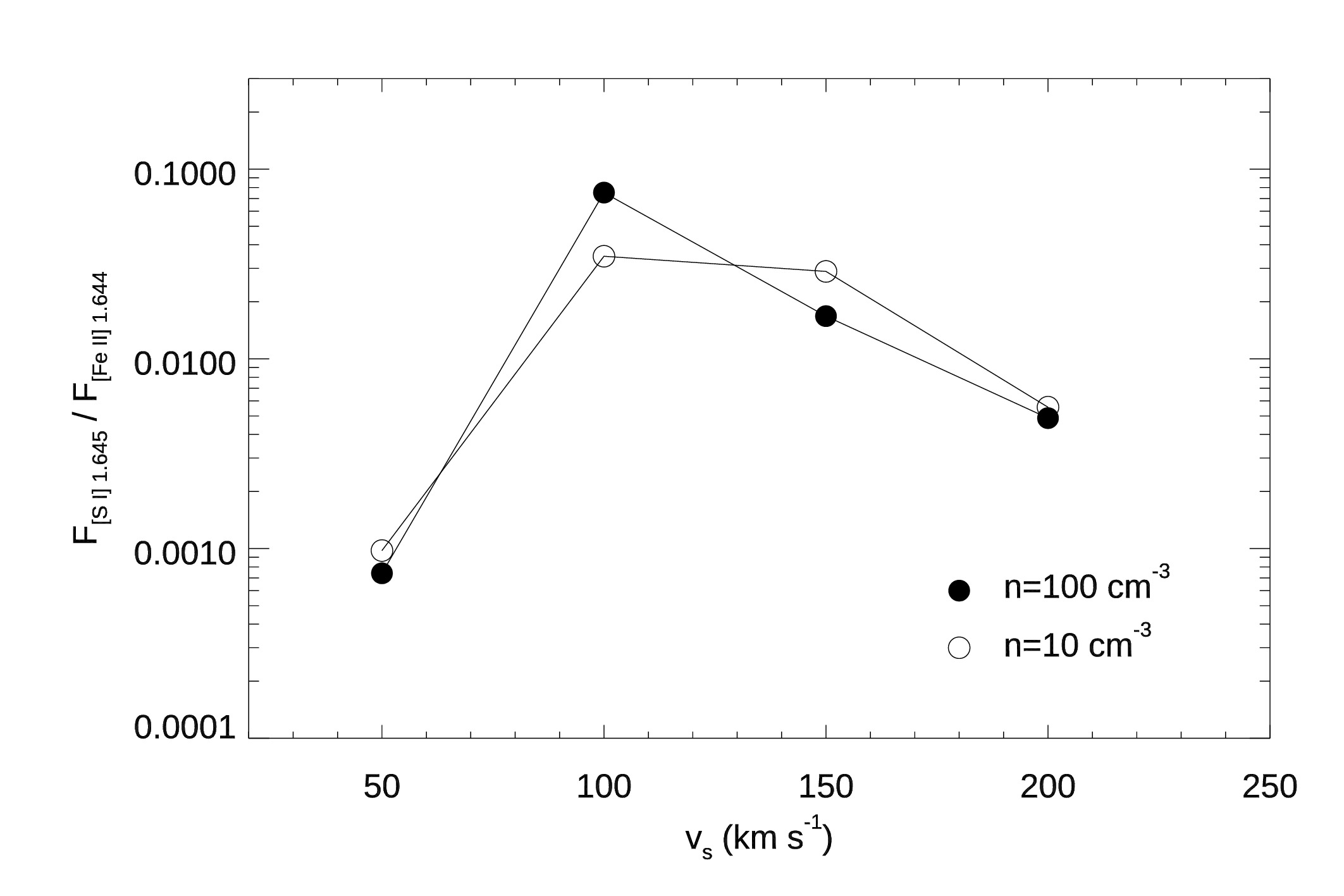}
\caption{The ratio of \sii\ 1.645 $\mu$m to \feii\ 1.644 $\mu$m line intensities 
as a function of shock speed $v_s$.  
Empty and filled symbols represent the cases for 
pre-shock ion density $n_0=10$ cm$^{-3}$ and 100 cm$^{-3}$, respectively. 
}
\label{fig_a1}
\end{figure}

FMKs are oxygen-rich SN ejecta material \citep[e.g.,][]{fesen01a,docenko10,morse04}.
The abundances of Si and Fe in FMKs are not known and 
they probably vary substantially among FMKs. 
As a representation, we adopt an abundance set where
H=[12.00], He=[11.00], C=[14.80], O=[16.00], 
Ne=[15.50], Mg=[14.80], Si=[14.70], 
S=[14.90], Ar=[14.50], Ca=[14.00], Fe=[14.70], and Ni=[14.00] in log scale.
These abundances are from \cite{morse04} except 
Fe, which we set equal to the Si abundance.  
Note that the abundances of O, S, Ar, and Ca are constrained by   
optical emission lines from shocked SN ejecta in Cas A, while 
the abundances of the other elements are not constrained \citep{morse04}.
We have run models 
with shock speed $v_s = 50$--200~\kms\ and pre-shock density 
$n_0 = 10$--$10^2$ cm$^{-3}$ using the 
shock code developed for SN ejecta \citep{raymond79,cox85,blair00,koo13}.
Magnetic field strength is fixed at 5 $\mu$G. 
The results are shown in Figure \ref{fig_a1}, where we see that the flux ratio $\fluxsii/\fluxfeii$ 
attains a maximum of 10\% at $v_s=100$~\kms. The ratio is less than a few percent 
at other velocities. Note that the result is for Si abundance equal to Fe abundance. 
If the Si abundance is greater than the Fe abundance by an order of magnitude, 
the intensity ratio can be $\sim 1$. 


%

\clearpage
\end{document}